\documentclass[referee]{raa} 

\usepackage{graphicx,times}             
\usepackage{natbib}
\usepackage{amssymb,amsmath}
\usepackage{amssymb,amsmath}
\usepackage{xspace}
\usepackage{enumitem} 
\usepackage{natbib}
\setlist[itemize]{label=\textbullet}
\bibpunct{(}{)}{;}{a}{}{,}
\usepackage[
    colorlinks=true,   
    linkcolor=blue,    
    citecolor=blue,    
    urlcolor=blue      
]{hyperref}
\usepackage{indentfirst}
\def\hi{\textsc{Hi}\xspace}
\begin{document}

  \title{A Statistical Study of HI Gas in AGN-Hosting and Satellite Galaxies from ALFALFA and FASHI
}

   \volnopage{Vol.0 (20xx) No.0, 000--000}      
   \setcounter{page}{1}          

   \author{Shuanghao Shu 
      \inst{1,2}
    \and Wenkai Hu
      \inst{1}
    \and Junqiang Ge
      \inst{3}
    \and Jinjiang Yu
      \inst{1,2} 
    \and Wenxiu Yang
      \inst{1,2}
   \and Furen Deng
      \inst{1,2}
    \and Yichao Li
      \inst{4}
   \and Yougang Wang
      \inst{1,2,4}
      \and Xuelei Chen
      \inst{1,2,4}
   }
\institute{State Key Laboratory of Radio Astronomy and Technology, National Astronomical Observatories, CAS, A20 Datun Road, Chaoyang District, Beijing, 100101, P. R. China; {\it wangyg@bao.ac.cn}\\
        \and
             School of Astronomy and Space Science, University of Chinese Academy of Sciences, Beijing 100049, China\\
        \and 
             National Astronomical Observatories, Chinese Academy of Sciences,
             Beijing 100012, China\\ 
        \and
            Key Laboratory of Cosmology and Astrophysics (Liaoning) \& College of Sciences, Northeastern University, Shenyang 110819, China\\
\vs\no
   {\small Received 20xx month day; accepted 20xx month day}}

\abstract{
We investigate the relative importance of Active Galactic Nucleus (AGN) feedback and environmental  processes using a large sample of \hi galaxies from the ALFALFA and FASHI surveys. By applying the optical spectroscopy from SDSS DR7/DR8 and the DESI survey, we analyse the gas content and physical properties of AGN-hosting galaxies in group environments. Our results show that AGN-hosting galaxies exhibit significantly suppressed star formation rates and \hi gas fraction, approximately one order of magnitude lower than star-forming counterparts, regardless of their group-centric position. AGN-hosting satellites exhibit a significant and persistent deficit in both gas fraction and SFR relative to normal satellites without AGN, even at the halo virial radius ($R/R_{180} \approx 1$). This suggests that cold gas depletion is primarily driven by internal AGN feedback before these galaxies experience intense environmental interactions. The relatively flat radial profiles of gas fraction and sSFR further indicate that the evolution of AGN-hosting satellites is governed by internal physical processes rather than environmental interactions. Moreover, the apparent increase in \hi gas at $R/R_{180} < 0.3$ is identified as an artifact of beam confusion. We conclude that for the AGN-hosting population, internal feedback is likely the prior quenching mechanism, while environmental effects act as a secondary, subsequent process.
}

   \authorrunning{S.-H. Shu, W.-K. Hu, J.-Q. Ge, J.-J. Yu, W.-X. Yang, F.-R. Deng, Y.-G. Wang,}            
   \titlerunning{ A Statistical Study of HI Gas in AGN-Hosting and Satellite Galaxies from ALFALFA and FASHI}  

   \maketitle

%
%
\section{Introduction}
\label{sect:intro}

Active galactic nuclei (AGN) are cosmic objects characterized by a compact, luminous core at the  center of a galaxy, powered by gas accretion onto a supermassive black hole, emitting tremendous energy across the entire electromagnetic spectrum. Early photographic surveys, notably by Carl Seyfert, identified galaxies with unusually broad emission lines originating from their luminous nuclei \citep{2020Univ....6..136F}. Subsequently, the development of radio astronomy and the 3C survey led to the discovery of quasi-stellar objects (quasars), linking these powerful radio sources to their optical counterparts \citep{1963Natur.197.1040S}.
In the contemporary understanding of galaxy formation and evolution, AGN play a crucial role through feedback mechanisms. AGN feedback is widely invoked to regulate star formation histories, particularly in massive galaxies, by driving the transition from blue, star-forming systems to red, quiescent ones \citep{2005Natur.433..604D,2009Natur.460..213C,2012ARA&A..50..455F,2022MNRAS.512.1052P}. Furthermore, this feedback can extend beyond the host galaxy, potentially influencing the evolution of satellite galaxies within the same dark matter halo \citep{2008MNRAS.388..587L}. However, recent observations and studies \citep{2013ApJ...774...66Z,2015ApJ...799...82C,2023A&A...678A.127V} also suggest that AGN activity can actually promote star formation rather than merely offsetting cooling and preventing the formation of new stars. Theoretical models \citep{2024ApJ...961L..39S} have also been established to provide a framework for understanding this phenomenon.


Hydrogen is the most abundant element in the universe, and neutral hydrogen (\hi) is the primary reservoir of cold gas in galaxies, serving as the fundamental fuel for both star formation and nuclear activity. Therefore, understanding the evolution of the \hi content is key to unraveling galaxy formation and evolution.  Numerous studies have examined the dependence of \hi mass (or \hi gas fraction) on stellar properties such as stellar mass, color (e.g., NUV-r), and star formation rate in nearby late-type/star-forming galaxies (e.g.,\citealt{2011ApJ...743...45E};\citealt{2011ApJ...732...93T};\citealt{2011ApJ...732...93T};\citealt{2012ApJ...756..113H}). However, the mechanisms that regulate this gas reservoir remain a subject of active debate. Two primary processes are thought to govern the removal or consumption of cold gas in satellite or central galaxies : (1)AGN feedback can heat or eject gas through radiation, jets, outflows \citep{2005Natur.433..604D,2006MNRAS.365...11C,2009Natur.460..213C,2012ARA&A..50..455F,2024Natur.632.1009W}, and (2) tidal interactions and ram-pressure stripping in dense environments, where gravitational forces and the intragroup/intracluster medium can directly strip gas from satellites \citep{1972ApJ...176....1G,1983ApJ...264...24M,1998ApJ...495..139M,2006MNRAS.369.1021M,2024A&A...686A.184D,2020A&A...638A.133L,2022MNRAS.516.4293W,2022A&ARv..30....3B}. While both mechanisms are known to suppress star formation, disentangling their relative importance in governing the cold gas content, particularly \hi, in satellite populations remains a challenge. Some studies point to the efficacy of ``indirect" AGN feedback \cite[e.g.][]{2017Natur.548..304P, 2019MNRAS.487.5889D}, while others emphasize the more direct and dominant role of environmental stripping \citep{2006MNRAS.369.1021M}.

Theoretical studies add further complexity. For example, \cite{2025ApJ...980..145L} showed that in the TNG simulations (The Next Generation Illustris Simulations; \citealt{2019ComAC...6....2N}), \hi content is primarily regulated by thermal-mode AGN feedback, whereas in the EAGLE simulations \citep{2015MNRAS.450.1937C}, stellar feedback dominates. Observations from the xGASS survey further suggest that in low-redshift galaxies, stellar feedback may play a more important role than AGN feedback \citep{2025ApJ...980..145L}. However, these simulation-based insights primarily focus on internal feedback mechanisms. In reality, galaxies are often subject to complex environmental processes. Some results \citep{2020MNRAS.493.1587H,2021PASA...38...35C} show that, beyond internal feedback, tidal forces and ram-pressure stripping serve as a crucial regulator of \hi content.
Consequently, it remains unclear from both observational and theoretical perspectives how the effects of AGN feedback and tidal forces compare in regulating the \hi content of satellite galaxies in the low-redshift universe, particularly within lower-mass groups where both mechanisms are likely to play a role.      

This study aims to directly compare the impact of these two mechanisms. We construct a large sample of galaxies by cross-matching \hi data from the Arecibo Legacy Fast ALFA Survey ( ALFALFA; \citealt{2018ApJ...861...49H}) and FAST All Sky HI Survey (FASHI; \citealt{2024SCPMA..6719511Z}) with optical spectroscopy from the Sloan Digital Sky Survey (SDSS; \citealt{2000AJ....120.1579Y}) and Dark Energy Spectroscopic Instrument (DESI; \citealt{2025arXiv250314745D}). We quantify both the AGN activity of central galaxies and the group/cluster related processes experienced by their satellites. By comparing the HI content of satellites orbiting central galaxies with different levels of AGN activity, as well as the HI content of galaxies under varying environmental effects, we aim to assess the relative roles of AGN feedback and group/cluster-related processes in regulating galaxy gas reservoirs.

This paper is organized as follows: after this Introduction, in Sec.\ref{sect:Obs}, we describe the ALFALFA and FASHI survey. Sec.\ref{sect:data} introduces the data products used in this study.  We present our main results in Sec.\ref{sect:analysis}.
Finally, we present the discussions and summarize this work in Sec.\ref{sect:summary and discussion}. In this paper, we adopt a flat $\Lambda$CDM cosmology with parameters: $\Omega_{\mathrm{m}} = 0.3$, $\Omega_{\Lambda} = 0.7$, and $
\rm H_{0} = 70~\mathrm{km~s^{-1}~Mpc^{-1}}$.

\section{Observations}
\label{sect:Obs}

\subsection{ALFALFA Survey}
\label{sect:alfalfa}

The \hi data used in this study are taken from the ALFALFA and FASHI. The ALFALFA survey \citep{2005AJ....130.2598G} is a large-scale blind \hi survey conducted with the Arecibo radio telescope. The survey was initiated in February 2005 and aimed to map approximately 7000 deg$^2$ of the high Galactic latitude sky with declination $\delta \lesssim +36^\circ$, systematically detecting \hi galaxies in the nearby universe. ALFALFA utilized the seven-beam L-band feed array (ALFA) installed on the 305\,m Arecibo telescope, observing in the 1335–1435\,MHz frequency range, corresponding to a redshift range of $z \lesssim 0.06$. The survey achieves an angular resolution of $\sim 3.5'$ and a spectral resolution of $\sim 5.5$\,km\,s$^{-1}$, providing high sensitivity and source identification capability.

The specific data used in this study are drawn from the $\alpha.100$ catalog of the ALFALFA survey released in 2018 \citep{2018ApJ...861...49H}, which covers approximately 7000\,deg$^2$ of the sky. The $\alpha.100$ catalog contains all sources detected upon completion of the survey, including a total of 31,502 \hi galaxies, among which 25,434 are high-confidence detections (code\,1) with good signal-to-noise ratios and reliable \hi parameter measurements, with the remaining 6,068 categorized as marginal detections (code 2). For each source, the catalog provides precise celestial coordinates, redshift, line widths ($W_{50}$ and $W_{20}$, representing the velocity widths at 50\% and 20\% of the peak flux), integrated flux, signal-to-noise ratio, and estimated \hi mass, with some sources cross-matched with SDSS and 2MASS data. 
In this study, we selected galaxies categorized under code 1 and code 2.

Because of the large sample size, uniform coverage, and high \hi sensitivity, the $\alpha.100$ catalog has been widely used in studies of the \hi mass function (HIMF) \citep{2020MNRAS.494.2090J}, low-mass galaxies \citep{2007PhDT........30S}, environmental effects \citep{2019MNRAS.490..566J}, and galaxy evolution \citep{2012ApJ...756..113H,2018ApJ...862...48A}.
In this work, we select galaxies from the $\alpha.100$ catalog that is successfully matched with the SDSS DR7 spectroscopic catalog as our primary sample, and further incorporate SDSS DR8 spectral and structural information to investigate the effects of AGN activity in central galaxies and tidal interactions on the \hi content of satellite galaxies.

\subsection{FASHI Survey}
\label{sect:fashi}

The FASHI \citep{2024SCPMA..6719511Z} survey aims to systematically cover the entire observable sky with declinations between $-14^\circ$ and $+66^\circ$, corresponding to an area of approximately 22,000\,deg$^2$, using the Five-hundred-meter Aperture Spherical radio Telescope (FAST) in China. The survey searches for the 21\,cm \hi emission line over the frequency range of 1050--1450\,MHz. The ultimate goal of the survey is to detect more than 100,000 \hi sources.  

As of June 2023, FASHI has completed observations over more than 7,600\,deg$^2$, corresponding to approximately 35\% of the FAST observable sky. The data used in this work are drawn from the catalog released in 2023, which contains a total of 41,174 extragalactic \hi sources, with an upper redshift limit of $z \lesssim 0.09$ and a frequency range of 1305.5-1419.5\,MHz. The covered sky region is $0^\mathrm{h} \leq \mathrm{RA} \leq 17.3^\mathrm{h}$, $22^\mathrm{h} \leq \mathrm{RA} \leq 24^\mathrm{h}$, $-6^\circ \leq \mathrm{DEC} \leq 0^\circ$, and $30^\circ \leq \mathrm{DEC} \leq 66^\circ$. 
The galaxies included in this first FASHI data release are predominantly located outside the footprint of the ALFALFA survey.
FASHI is a blind extragalactic survey conducted in drift-scan mode. The median detection sensitivity is approximately 0.76\,mJy\,beam$^{-1}$, with a velocity resolution of $\sim 6.4$\,km\,s$^{-1}$ at 1.4\,GHz. The corresponding integration time is 21s. The spatial resolution and sky coverage surpass those of previous large-scale \hi surveys.  
As summarized in Table1 \citep{2018ApJ...861...49H}, the ALFALFA survey has a typical system temperature ($T_{\rm sys}$) of 26–30 K and a gain of $11\,\rm K\,Jy^{-1}$ for its central beam. In contrast, the FASHI survey \citep{2024SCPMA..6719511Z} achieves a much lower $T_{\rm sys}$ of 16–19 K and a significantly higher gain of $13\text{--}17\,\rm K\,Jy^{-1}$. FASHI's map median sensitivity is substantially deeper than ALFALFA’s map rms of $1.86\,\rm mJy\,beam^{-1}$ at a $10\,\rm km\,s^{-1}$ resolution.  98.3\% of FASHI sources having better SNR compared to ALFALFA sources, the latter will have significantly lower SNR than the faint sources in FASHI. Furthermore, FASHI's depth enables a more complete census of low-mass galaxies; while ALFALFA may underestimate the population below $10^9\,M_\odot$ because of its sensitivity limits and cosmic variance\citep{2022ApJ...941...48L}, FASHI has already demonstrated superior detection rates in the $10^8\text{--}10^9\,M_\odot$ range\citep{2024SCPMA..6719511Z}.
While FASHI is based on a single pass survey, ALFALFA was conducted
using two separate scans. The absence of ALFALFA
sources in the redshift range $0.06 < z < 0.09$ is due to local radar interference at
the observatory.

With its high sensitivity, wide frequency coverage, and large sky area, FASHI represents the most prominent blind \hi survey following ALFALFA, providing a crucial observational basis for understanding the distribution of \hi gas in the local universe, the mechanisms of galaxy formation and evolution, and the large-scale structure of the Universe.

\subsection{The Sloan Digital Sky Survey}
\label{sect:sdss}
In this study, the optical spectroscopic galaxy sample is drawn from the SDSS \citep{2000AJ....120.1579Y}. We adopt the galaxy group catalog constructed by \citet{2007ApJ...671..153Y} based on the SDSS DR7 main galaxy sample. This catalog employs a halo-based group finder to classify $639,359$ galaxies into groups, derive the dark matter halo mass associated with each galaxy, and provide central/satellite classifications. Owing to its detailed environmental information, this group catalog is particularly well-suited for investigating the influence of large-scale structure, group environment, and central/satellite properties on galaxy evolution.

To further obtain the internal physical properties of galaxies, such as spectroscopic redshift, optical emission-line fluxes, stellar mass ($M_\star$), star formation rate (SFR), and gas-phase metallicity ($12+\log(\mathrm{O/H})$), we also employ the processed SDSS DR8 spectroscopic galaxy catalog provided by the MPA-JHU group.
The MPA-JHU catalog is based on SDSS spectroscopic data, with detailed descriptions of the data reduction and analysis given in a series of papers (\citealt{2003MNRAS.346.1055K};\citealt{2004MNRAS.351.1151B};\citealt{2004ApJ...613..898T};\citealt{2007ApJS..173..267S}). By applying spectral synthesis and emission-line measurements, this catalog derives key physical properties for a large sample of galaxies and has become one of the most widely used datasets for studies of galaxy formation and evolution.

These two catalogs are complementary in their functions: the Yang catalog provides environmental classifications and halo mass information, making it suitable for analyses of group structures and central/satellite dependencies; while the MPA-JHU catalog offers spectroscopically derived galaxy physical properties, enabling us to link environmental characteristics with internal properties. Together, they allow for a systematic investigation of AGN activity, stellar feedback, and gas evolution processes across different environments.

\subsection{DESI}
\label{sect:desi}
We utilize the DESI DR1 \citep{desicollaboration2025datarelease1dark} AGN/QSO value-added catalog (VAC; Juneau et al., in prep.) to identify active galactic nuclei (AGN) within the FASHI sample. We cross-match our sources with the DESI DR1 AGN/QSO VAC.
Each spectrum in this catalog is assigned a type (STAR, GALAXY, or QSO) and a redshift estimate using the Redrock pipeline \citep{2023AJ....165..144G}. For AGN classification, we adopt the optical emission-line parameters provided in the VAC through FastSpecFit v2.1 \citep{2023ascl.soft08005M}, and follow the same optical and ultraviolet diagnostic diagrams (e.g., the Baldwin-Phillips-Terlevich (BPT) diagnostic diagram\citep{1981PASP...93....5B}) applied therein. In addition, the catalog incorporates AGN classifications derived from WISE mid-infrared data, which are particularly useful for identifying dust-obscured AGN \citep{2016ascl.soft04008L}. The DESI spectrographs achieve a spectral resolution of $R \approx 2000\text{--}5000$ across a broad wavelength range of 360–980 nm, facilitated by a three-channel design using dichroic splitters. ALFALFA and FASHI have an angular resolution of $3.5'$ and $2.95'$, respectively. During the cross-matching, we employ a spatial matching radius of $3.5\arcmin$ and further require consistent redshifts to select a reliable AGN sample.

By cross-matching with the DESI AGN catalog, we identified optical AGN within the FASHI sample, specifically some sources that lacked classification in the MPA-JHU catalog. In particular, we made use of the \texttt{OPT\_UV\_TYPE} field in the DESI DR1 AGN/QSO value-added catalog, which encodes [N\,\textsc{ii}]-BPT diagnostic results using a bitmask scheme. This allows us to assign spectral types, redshifts, emission-line parameters, and AGN classification labels to a subset of FASHI galaxies, thereby providing the necessary basis for investigating the connection between H\,\textsc{i} gas content and AGN activity.

\section{Sample and Data reduction}
\label{sect:data}
\subsection{Spectral Classification}
We cross-matched the ALFALFA $\alpha.100$ catalog \citep{2018ApJ...861...49H} with the SDSS DR7 spectroscopic sample \citep{2007ApJ...671..153Y}. A positional matching radius of $3.5'$ was adopted, and we further required redshift consistency between the optical and \hi measurements ($|z_{\mathrm{SDSS}} - z_{\mathrm{HI}}| < 0.001$) ,  selecting the nearest optical counterpart in cases of multiple matches. This yielded a total of 19,063 \hi sources with available spectra, hereafter referred to as the $\alpha.100$--SDSS sample. Our sample objects can all be
matched to corresponding sources in the ALFALFA-SDSS Galaxy catalog\citep{2020AJ....160..271D}. The number of objects in our sample is smaller than that in the ALFALFA-SDSS catalog. The difference is that the official ALFALFA-SDSS Galaxy catalog also includes counterparts identified via photometry for low-S/N sources (Code 2), in addition to using morphology and visual inspection to handle ambiguous or uncertain optical identifications. In the FASHI matching strategy, the \hi catalog was first cross-matched with the Siena Galaxy Atlas (SGA,  \citealt{2023ApJS..269....3M}), yielding 14,072($\approx33.7\%$) optical counterparts, and the redshifts provided by the SGA were directly adopted for these matches. For the remaining 27,669 \hi sources without SGA counterparts, we further cross-matched with the SDSS DR7 spectroscopic catalog, resulting in 2,900($\approx7.0\%$) galaxies with available spectra. Remaining 10,975 sources in FASHI ($\approx26.3\%$) have only photometric redshifts, and the remaining 13,794 sources ($\approx 33.0\%$) lack identifiable optical counterparts due to their location near the Galactic plane or contamination by Galactic stars. It should be noted that many SGA-matched galaxies also have SDSS spectra, which were not further exploited in the original FASHI catalog. In this work, we further supplemented the FASHI sample by cross-matching all SGA-matched galaxies with the SDSS DR7/DR8 spectroscopic data using the same strategy. This procedure expanded the sample of \hi galaxies with complete spectroscopic properties to 13,763 sources, which we refer to as the FASHI--SDSS sample.

To select AGN, we performed spectral classification of all emission-line galaxies using the BPT diagram, which separates AGN, star-forming galaxies, and composite galaxies. 
We first required that the fluxes of H$\alpha$, H$\beta$, [O\,\textsc{iii}], and [N\,\textsc{ii}] have signal-to-noise ratios greater than \textbf{$3$}. Galaxies are classified based on their position in the BPT diagram (using emission line ratios: [O\,\textsc{iii}]/H$\beta$ vs. [N\,\textsc{ii}]/H$\alpha$). Galaxies located below the \cite{2003MNRAS.346.1055K} (hereafter Ka03) line are classified as star-forming, those above the \cite{2001ApJS..132...37K} (hereafter Ke01) line are classified as narrow-line AGN, and galaxies between these two lines are classified as composites (starburst + AGN) \citep{2006MNRAS.372..961K,1998A&AS..132..181W}. Specifically, AGN are identified if they satisfy:
\begin{align}
\log_{10}\left(\frac{\rm [O\,III]}{\rm H\beta}\right) &\geq \frac{0.61}{\log_{10}([{\rm N\,II}]/\rm H\alpha) - 0.05} + 1.3, \\
\log_{10}\left(\frac{\rm [O\,III]}{\rm H\beta}\right) &\geq \frac{0.61}{\log_{10}([{\rm N\,II}]/\rm H\alpha) - 0.47} + 1.19,
\label{AGN selection}
\end{align}
where $\text{[O III]} \lambda 5007$ and $\text{[N II]} \lambda 6584$ are the forbidden emission lines of doubly ionized oxygen and singly ionized nitrogen, respectively; $\text{H}\alpha$ and $\text{H}\beta$ represent the hydrogen recombination lines of the Balmer series. 

In this study, we focus on AGN hosts identified via optical emission-line diagnostics. We cross-matched the ALFALFA $\alpha.100$ and FASHI \hi catalogs with the SDSS DR8 spectroscopic database. For the matched galaxies, we adopted the BPT classifications provided by the MPA--JHU catalog   \citep{2004MNRAS.351.1151B}. This catalog encodes classifications as: $-1$ (Unclassifiable), $1$ (Star-forming galaxies), $2$ (Low S/N star-forming galaxies), $3$ (Composite galaxies), $4$ (AGN, excluding the Low-Ionization Nuclear Emission-line Regions(LINERs)), $5$ (Low S/N LINERs).   


The Ka03 \citep{2003MNRAS.346.1055K} line is an empirical division defined to separate star-forming galaxies from those Seyfert–HII composite objects, whose emission lines reflect the combined influence of both star formation and AGN activity. Ke01 \citep{2001ApJS..132...37K} line used a more stringent cut by combining stellar population synthesis (PEGASE 2.0, \citealt{1997A&A...326..950F}) and photoionization (MAPPINGs v3.0) models to establish a theoretical maximum starburst line. Galaxies lying above this line are likely to be dominated by an AGN. In this study, by adopting the Ka03 criterion, we ensure our sample encompasses a broad range of active galaxies, including both Seyferts and LINERs, as well as composite systems. In the Galaxy Properties for DR8 spectra from MPA--JHU catalog, the \texttt{galSpecExtra} table provides emission-line classifications based on the BPT diagram, following the methodology of \cite{2004MNRAS.351.1151B}.

As shown in Table~\ref{BPT number}, this yielded  789 AGN and 569 LINERs in the $\alpha.100$--SDSS sample, and 535 AGN and 497 LINERs in the FASHI--SDSS sample. The total number of AGN hosts (including LINERs) we identify from the ALFALFA  $\alpha.100$ catalog is 1358. This is smaller than the the $\sim$1560–1600 AGN reported in previous studies \citep{2019MNRAS.482.5694E,2022ApJ...933L..12G}. We attribute this difference primarily to methodology: those works included galaxies without individual \hi detections by employing spectral stacking, whereas our analysis relies solely on systems with direct ALFALFA \hi measurements.

To expand our sample, we further cross-matched the remaining FASHI galaxies with the DESI DR1 AGN/QSO value-added catalog (VAC), using the [N\,\textsc{ii}]-BPT diagnostic encoded in the \texttt{OPT\_UV\_TYPE} field. This cross-match identified an additional 43 AGN and 59 LINERs.

\begin{figure}
\centering
\includegraphics[width=\textwidth]{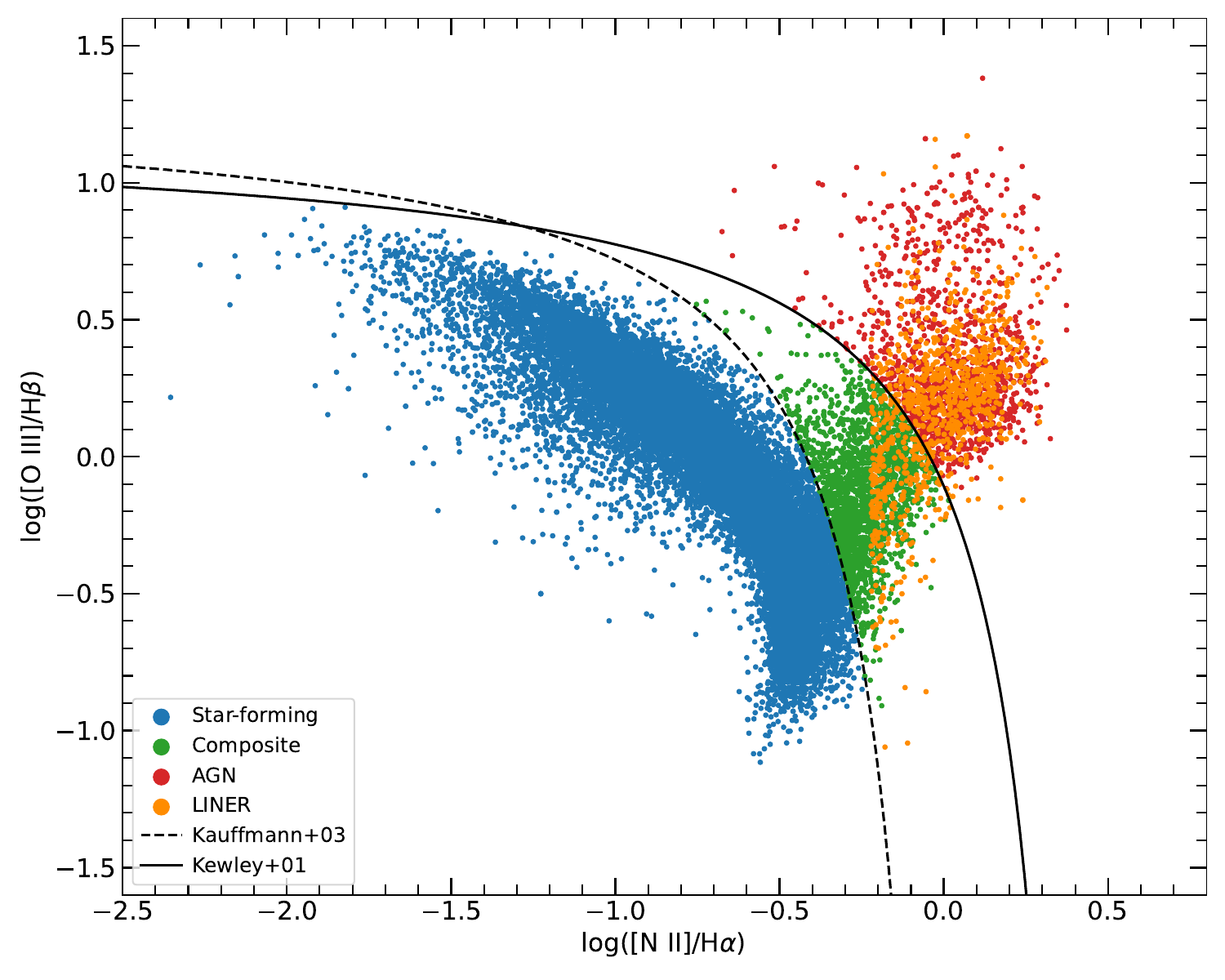}
\caption{The BPT diagnostic diagram for the sample galaxies. The vertical axis shows the $\log([\text{O III}]/\text{H}\beta)$ ratio, while the horizontal axis shows $\log([\text{N II}]/\text{H}\alpha)$. The dashed line represents the Ka03 empirical division, separating pure star-forming galaxies (blue) from those with potential AGN contribution. The solid line represents the Ke01 theoretical upper limit for star-forming galaxies; objects above this line are classified as narrow-line AGN (red/orange). Galaxies located between the Ka03 and Ke01 lines are identified as composite systems (green), exhibiting both star formation and AGN activity.
} 
\label{fig:BPT_dia}
\end{figure}

Figure~\ref{fig:BPT_dia} shows the location of our combined sample on the [N\,\textsc{ii}]-BPT diagram, where star-forming, composite, AGN, and LINERs galaxies are color-coded blue, green, red, and yellow, respectively. 
Low-S/N star-forming galaxies were excluded from the analysis. The emission-line ratio of $\rm [O\,III]/{\rm H\beta}$ is highly sensitive to the hardness of the ionization source. AGN radiation is harder than stellar radiation and can more efficiently produce ionized ions. Consequently, the ratio is typically higher in AGN. The ratio of $\rm [N\,II]/{\rm H\alpha}$ is sensitive to the gas metallicity and the ionization parameter. In AGN or metal-rich star-forming regions, the abundance of nitrogen is relatively high, and the radiation field of an AGN can effectively produce $\rm [N\,II]$.
We restrict our analysis to galaxies that host an AGN, without drawing a sharp distinction between Seyferts and LINERs. While Ke01 proposes a more stringent cut, we are interested in the statistical ensemble and therefore do not apply this stricter classification at present.

\begin{table}[ht]
\centering
\begin{tabular}{|c|c|c|c|c|c|}
\hline
Sample & Star-Forming& Lowe-S/N SF & Composite &AGN & LINERs \\
\hline
$\rm \alpha.100-SDSS$ & 13280 & 2820 & 1269&789&569 \\
\hline
FASHI-SDSS& 9455 & 2165& 875&535&497 \\
\hline
FASHI-DESI&1944&0&125&43&59\\
\hline
\end{tabular}
\caption{BPT class for all samples.}
\label{BPT number}
\end{table}

\subsection{Central and Satellite Galaxies}
\label{sect:satellite}
In this study, we cross-matched the \hi galaxies from ALFALFA and FASHI 
with the group catalog of \cite{2007ApJ...671..153Y}. 
Since the catalog of \cite{2007ApJ...671..153Y} is constructed from a subset of SDSS DR7 \citep{2005AJ....129.2562B}, we utilized the (\texttt{plate, mjd, fiberid}) identifiers to perform an exact match between the SDSS DR7 galaxy catalog and the Yang catalog. Through this matching, we obtained the corresponding \texttt{VAGC\_ID} for each galaxy. This identification allowed us to directly extract group membership information—specifically the classification of central and satellite galaxies—from the Yang catalog. To ensure sample completeness, we restricted our analysis to galaxies in sectors with $f_{\mathrm{gotmain}} \geq 0.7$, and adopted Sample 1, which exclusively consists of galaxies with SDSS spectroscopic redshifts.



For the galaxies not matched in the first step, we extracted their coordinates ($\rm R.A.$, $\rm Decl.$) and spectroscopic redshifts from the ALFALFA and FASHI catalogs. We then cross-matched these sources against the galaxies in the \cite{2007ApJ...671..153Y} group catalog, using a matching radius of 3.5 arcmin and a redshift tolerance of $\Delta z < 0.001$ to minimize spurious matches. In cases with multiple counterparts, we preferentially selected the nearest neighbor consistent in both position and redshift, and excluded sources with ambiguous matches or large redshift discrepancies.


After matching, each \hi galaxy was assigned a group ID together with its hierarchy properties within the group. The \cite{2007ApJ...671..153Y} catalog further identifies, for each group, the brightest galaxy and the most massive galaxy. In this study, we define a galaxy as a \textit{central galaxy} if it possesses a valid group ID and is identified by the catalog as the most massive galaxy of its group. All remaining group members are classified as \textit{satellite galaxies}. \hi galaxies lacking a group association (group ID = 0) are treated as \textit{field galaxies} and are excluded from the central/satellite analysis.

Specifically, for the $\alpha.100$--SDSS \hi sample containing 19,063 galaxies, 15,897 sources were successfully matched to the SDSS group catalog in the first step, yielding 12,342 central galaxies and 3,555 satellite galaxies. An additional coordinate- and redshift-based cross-matching step recovered 578 central galaxies and 204 satellite galaxies, bringing the total number of matched galaxies to 16, 679, which corresponds to 87.50\% of the original ALFALFA sample. Similarly, for the FASHI-SDSS sample of 13,763 galaxies, 11,320 were initially matched (9,027 central galaxies and 2,293 satellite galaxies). The subsequent coordinate-redshift match added 579 central galaxies and 128 satellite galaxies, resulting in a total of 12,027 matched galaxies (87.39\% completeness). For the FASHI-DESI sample of 2,183 galaxies, the coordinate-redshift match contributed 41 central galaxies and 6 satellite galaxies.

This comprehensive matching strategy ensures that the majority of \hi-detected galaxies are properly classified within their group environments, providing a reliable basis for investigating how AGN activity and environmental processes (such as tidal interactions) influence the neutral gas content of galaxies.

\begin{table}[ht]
\centering
\begin{tabular}{|c|c|c|c|c|}
\hline
Sample & $\rm Central^1$&  $\rm Central^2$&$\rm Satellite^1$&$\rm Satellite^2$ \\
\hline
$\rm \alpha.100-SDSS$ & 12342&578 & 3555 & 204 \\
\hline
FASHI-SDSS& 9027 & 579&2293&128 \\
\hline
FASHI-DESI&0&41&0&6\\
\hline
\end{tabular}
\caption{Central and satellite galaxies in the entire sample. 1: Matches satellite galaxies to central galaxies via an exact match of plate, mjd, fiberid, and the corresponding $\rm VAGC\_ID$.
2: Represents an additional matching method that also uses celestial coordinates and spectroscopic redshift.}
\label{central_satellite number}
\end{table}

\subsection{Galaxy properties}
\label{sect:properties}
We consider the following several properties for  basic analysis of the $\rm \alpha.100-SDSS$, FASHI-SDSS, and FASHI-DESI HI samples.  
\begin{itemize}[leftmargin=0.4cm]
\item $\boldsymbol{\rm M}_{\mathrm{HI}}$: the neutral hydrogen mass of galaxies, obtained from the ALFALFA and FASHI catalogs. We compute the \hi mass following \citep{2018ApJ...861...49H}:
\begin{equation}
 \rm M_{\rm HI}= 2.356 \times 10^5 \; \left(\frac{D_L}{\rm Mpc}\right)^2 \; \left(\frac{S_{\rm HI}}{\rm Jy\,km/s}\right)
\label{MHI}
\end{equation}

\item 
\noindent
$\boldsymbol{\rm M}_{\ast}$: Stellar mass. In this work, we adopt the \texttt{lgm\_tot\_p50} parameter from the MPA--JHU DR8 catalog as the indicator of the total stellar mass. This parameter represents the median of the posterior distribution of stellar mass, 
derived from Bayesian SED fitting based on the SDSS total galaxy photometry (\texttt{model magnitudes}) and the stellar population synthesis library of \citet{2003MNRAS.344.1000B}:
\[
\log\left(\frac{M_\ast}{M_\odot}\right).
\]
Specifically, the SDSS five-band \texttt{model magnitudes}, which trace the integrated light distribution of galaxies, are used to constrain colors and luminosities. These are then compared with theoretical SEDs generated from assumed star formation histories (SFH), metallicities, and dust attenuation parameters. The observed photometry is fitted using a Bayesian framework, yielding a posterior distribution of stellar mass. The median value (p50) is adopted as the best estimate, while the 16th and 84th percentiles (p16 and p84) are provided as uncertainty estimates.

\item \noindent
$\boldsymbol{\log \mu_\ast}$: Stellar surface mass density, defined as
\[
\log \mu_\ast = \log \left( \frac{M_\ast}{2 \pi R_{50,z}^2} \right),
\]
where $R_{50,z}$ is the half-light radius in the SDSS $z$ band. The parameter can be obtained by cross-matching the MPA--JHU DR8 galaxy catalog with either the SDSS \texttt{PhotoObj} table (using \texttt{plate--MJD--fiber} or \texttt{objID}) or the NYU--VAGC catalog \citep{2005AJ....129.2562B}.

\item \noindent
$\boldsymbol{\rm SFR}$: Star formation rate. In this study, we adopt the total SFR estimate \texttt{SFR\_tot\_p50} from the MPA--JHU DR8 galaxy catalog as the primary indicator of the overall SFR. 
Star formation rates (SFRs) are derived within the SDSS fiber aperture from nebular emission lines following the methodology of \citet{2004MNRAS.351.1151B}. The contribution outside the fiber is estimated using broadband photometry as described in \citet{2007ApJS..173..267S}. For AGN hosts and galaxies with weak emission lines, SFRs are instead inferred from photometric indicators.
The parameter \texttt{SFR\_tot\_p50} corresponds to the median of the posterior probability distribution of the total SFR,
\[
\log \left( \mathrm{SFR} \,/\, M_\odot \,\mathrm{yr}^{-1} \right),
\]

\item $\boldsymbol{\rm sSFR}$: specific star-formation rate, defined as the ratio of star-formation rate to stellar mass:
\[
\mathrm{sSFR} = \frac{\mathrm{SFR}}{M_\ast}.
\]
In this work, sSFR is derived by combining the likelihood distributions of both star-formation rate and stellar mass. We adopt the fiber sSFR median (50th percentile) provided by the MPA–JHU DR8 catalog as our primary indicator. In addition, the total sSFR median is used as a cross-check of the star-formation efficiency for the entire galaxy.

\item 
$\boldsymbol {\rm u-r}$: The $\rm u-r$ color index is defined as the difference between the dereddened SDSS $u$-band and $r$-band magnitudes:
\[
\rm u-r = m_u - m_r.
\]
This parameter provides a diagnostic of the stellar population. Galaxies with lower $\rm u-r$ values appear bluer and are typically dominated by ongoing star formation, while those with higher $\rm u-r$ values appear redder and are dominated by older stellar populations. The photometric data are taken from the SDSS \texttt{PhotoObj} table and the NYU–VAGC catalog.

\item $\boldsymbol{\rm log(O/H) + 12}$: The gas-phase metallicity is a measure of the relative abundance of oxygen in the ionized gas at the center of galaxies. The values are derived from the MPA--JHU DR8 catalog, based on spectral fitting with the models of \cite{2001MNRAS.323..887C} and \cite{2004ApJ...613..898T}.

\item $\boldsymbol{\rm \sigma_{\ast}}$: logarithm of central stellar velocity dispersion, taken from the MPA--JHU DR8 catalog.

\end{itemize}

As mentioned above, $\mu_\ast$ and $\rm u-r$ are known to correlate tightly with the \hi mass fraction, as demonstrated by some scaling relation studies \citep{2004ApJ...611L..89K,2009MNRAS.397.1243Z,2018MNRAS.476..875C,2025ApJ...980..145L}. Other galaxy properties—such as the SFR, sSFR, $\rm M_\ast$, ${\rm log(O/H) + 12}$ and $\rm \sigma_{\ast}$ will be combined with the $\rm M_{\mathrm{HI}}$ and \hi gas fraction ($M_{\mathrm{HI}}/M_\ast$) to explore how AGN strength correlates with gas depletion or retention. For the FASHI-DESI sample, we cross-matched the DESI AGN/QSO Value-Added Catalog
with the DESI DR1 Galaxies value-added catalog
\citep{2020yCat..22420008Z,2024A&A...691A.308S,2024ApJ...961..173Z,2025arXiv250314745D}, which provides estimates of the 
$\rm M_*$, SFR,  $\sigma_\ast$, $D_{N}4000$ and Half-light radius for all DESI DR1 galaxies with reliable redshift determinations(defined by SPECTYPE $==$ GALAXY and ZWARN $==$ 0). 
Stellar masses are derived using the SED fitting code CIGALE \citep{2019A&A...622A.103B}, while additional stellar population parameters (e.g., age and metallicity) for the DESI Bright Galaxy Sample (BGS) are obtained using STARLIGHT \citep{2005MNRAS.358..363C}  based on DESI optical spectra.
Emission line fluxes are measured from the rest-frame spectra.
However, this catalog does not include other physical quantities commonly available
in the SDSS MPA–JHU database, such as 
gas-phase metallicity ($12+\log(\mathrm{O/H})$) or $\rm u-r$.

To quantify the influence of AGN activity on the cold gas reservoirs of galaxies, additional physical parameters can be employed as tracers of AGN activity. The most direct indicator is the extinction-corrected [O\,\textsc{iii}] $\lambda5007$ emission-line luminosity ($L_{\mathrm{[O\,III]}}$) \citep{2003MNRAS.341...33K,2008ApJ...679...86W}, which traces the ionizing power of the central engine and is widely used as a proxy for the AGN bolometric luminosity \citep{2004ApJ...613..109H}.    
The ratio of [O\,\textsc{iii}] to H$\beta$ ($\mathrm{[O\,III]}/\mathrm{H}\beta$), being sensitive to the hardness of the ionizing radiation field, thus serves as another useful tracer of AGN activity. 
Together, these parameters allow a multi-dimensional characterization of AGN feedback effects, linking nuclear activity to the global gas content and star formation properties of galaxies.

%

\section{Results}
\label{sect:analysis}

\subsection{Properties of \hi galaxies in the sample}
\label{Properties}
In this study, we construct two \hi-selected galaxy samples by cross-matching the \textsc{ALFALFA} 100\% catalog with the SDSS DR7,
yielding a total of 19,063 \hi sources, and by cross-matching the \textsc{FASHI} survey 
with SDSS DR7/DR8, yielding an additional 13,763 \hi sources. 
In our matched sample, FASHI includes 1,449 galaxies with $z > 0.06$ (accounting for 7.7\% of the total), 160 of which are AGN hosts. Additionally, we identified five galaxies with low HI masses ($M_{HI} < 10^{6.5} M_\odot$).
Within these samples, galaxies are classified into star-forming (SF), active galactic nuclei (AGN), composite, and unclassified systems based on their optical emission-line diagnostics. 
We present the basic properties of these \hi-detected galaxies, 
including their SNR, redshift, integrated \hi flux, Hubble distance, \hi mass, and angular distances relative to their optical counterparts, to explore the role of AGN activity and star formation in shaping the 
\hi properties of galaxies in the local Universe.

\medskip
\noindent
\textbf{Signal-to-noise ratio.}  
The signal-to-noise ratios ($SNR$) of the \hi detections were taken directly from the ALFALFA and FASHI catalogs. 

\medskip
\noindent
\textbf{Redshift distribution.}  
The \hi-selected galaxies are primarily at low redshift. Those from ALFALFA have $z \lesssim 0.06$, the FASHI sample extends this range to $z \sim 0.09$.

\medskip
\noindent
\textbf{\hi mass and integrated flux.}  
The \hi mass distribution spans approximately $10^{5.67} - 10^{10.93}\,M_\odot$. 
The integrated flux of the present \hi detections was all taken from those provided in the ALFALFA and FASHI catalogues.

\medskip
\noindent
\textbf{Optical counterparts.}  
Cross-matching with the optical galaxy catalogs shows that \hi galaxies have counterparts within an angular separation of $<3.5$ arcmin from the \hi centroid, consistent with the standard matching criteria used in large-area \hi surveys such as ALFALFA and FASHI. 

\begin{figure}
\centering
\includegraphics[width=\textwidth]{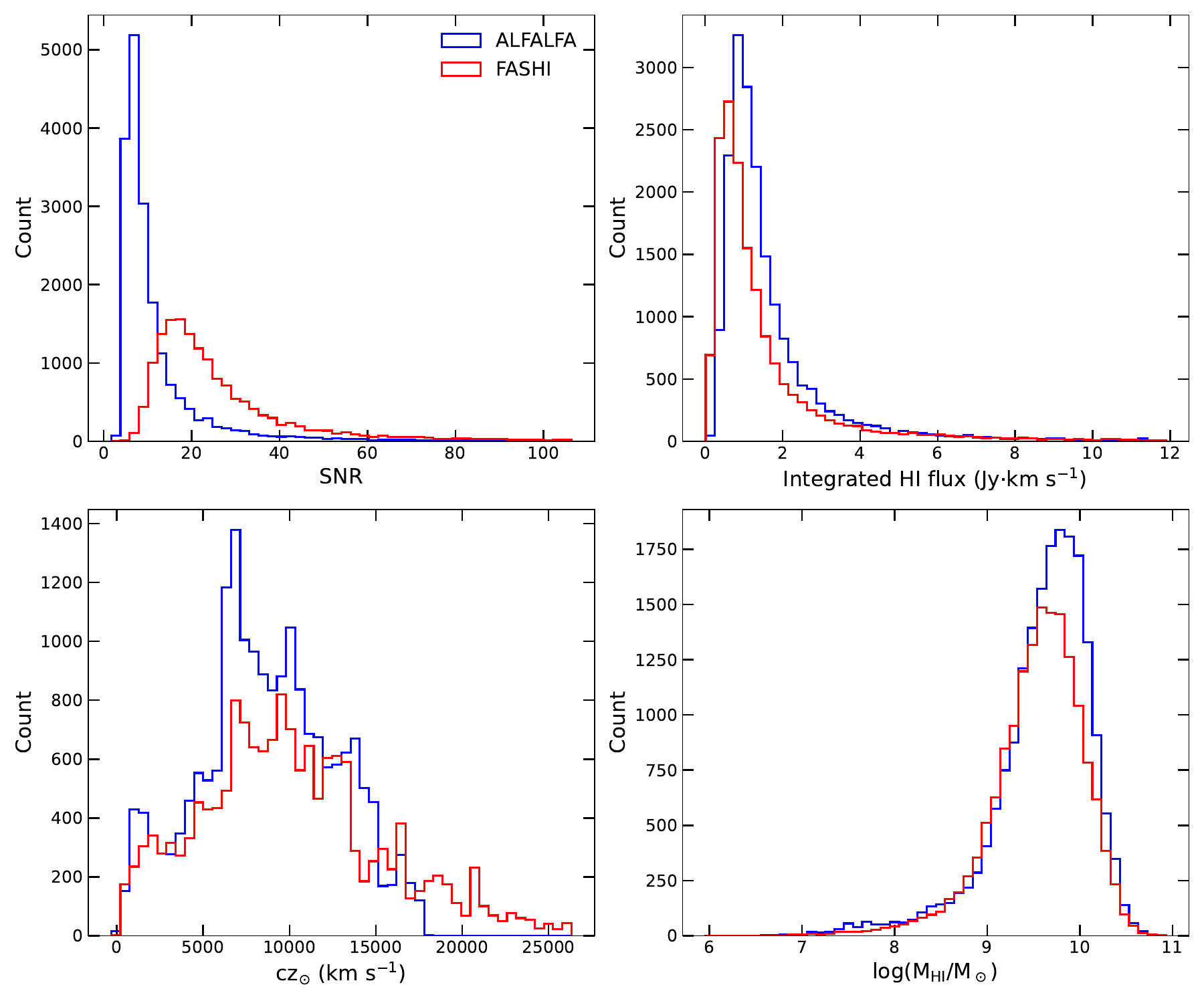}
\caption{Histograms of the SNR (top left), integrated \hi flux (top right), heliocentric velocity (bottom left), and \hi mass of the \hi galaxy (bottom right), together with the corresponding galaxy counts. Blue and red represent the ALFALFA samples and the FASHI samples, respectively.} 
\label{fig:par_distribution}
\end{figure}

\begin{figure}
\centering
\includegraphics[width=0.5\textwidth]{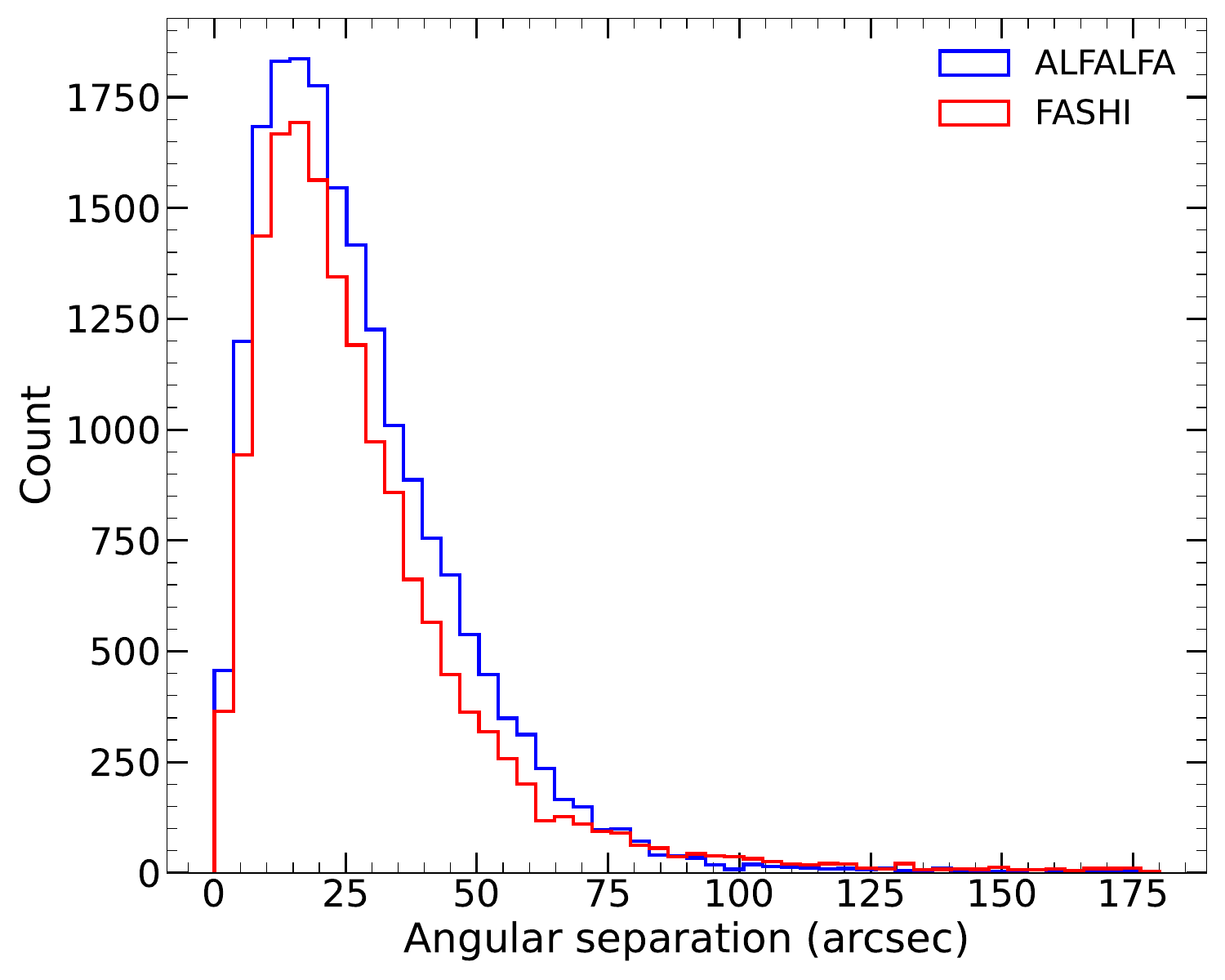}
\caption{Histogram of the angular separations between \hi galaxies and their optical counterparts, together with the corresponding galaxy counts.
Blue and red represent the ALFALFA samples and the FASHI samples, respectively.} 
\label{fig:sep_distribution}
\end{figure}

Figure \ref{fig:par_distribution} presents the parameter distributions of the \hi galaxies with optical counterparts in the two catalogues we selected. 
In ALFALFA, the SNR spans 2.4–300, while in FASHI it extends from 5.0 to 3,300.  
The integrated \hi flux range is 0.11–43.41 $\mathrm{Jy\cdot km\,s^{-1}}$ for ALFALFA, whereas FASHI covers 0.015-833.3 $\mathrm{Jy\cdot km\,s^{-1}}$, demonstrating the substantially deeper flux limit achieved by FAST’s single-dish sensitivity.
ALFALFA spans redshifts up to $z=0.06$, whereas currently available FASHI sample pushes the limit to $z=0.09$. The \hi mass distribution covers $10^{5.67}\,M_{\odot}$--$10^{11}\,M_{\odot}$. As shown in Figure \ref{fig:sep_distribution}, we compute the angular separation between each \hi galaxy and its optical counterpart. The Histogram shows the distribution of angular separations between \hi galaxies and their matched optical counterparts. We find that the vast majority of objects exhibit separations smaller than 100 arcsec.

\subsection{The HI Gas of AGN-Hosting Galaxies}

\begin{figure}
\centering
\includegraphics[width=\textwidth]{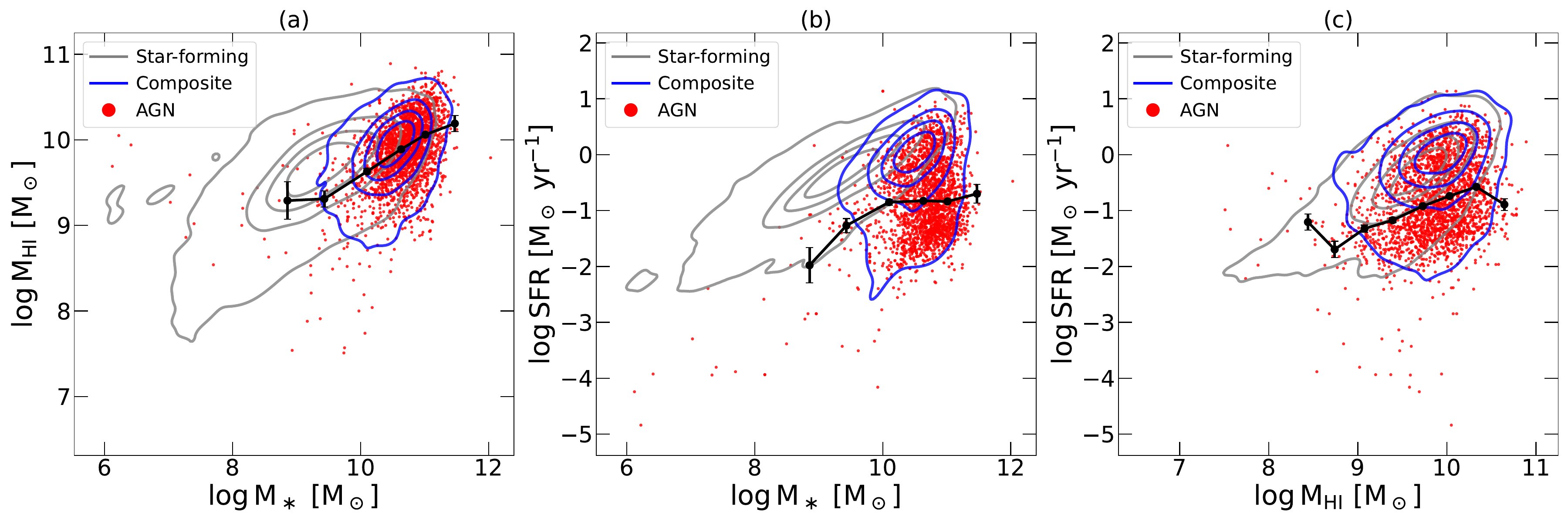}
\includegraphics[width=\textwidth]{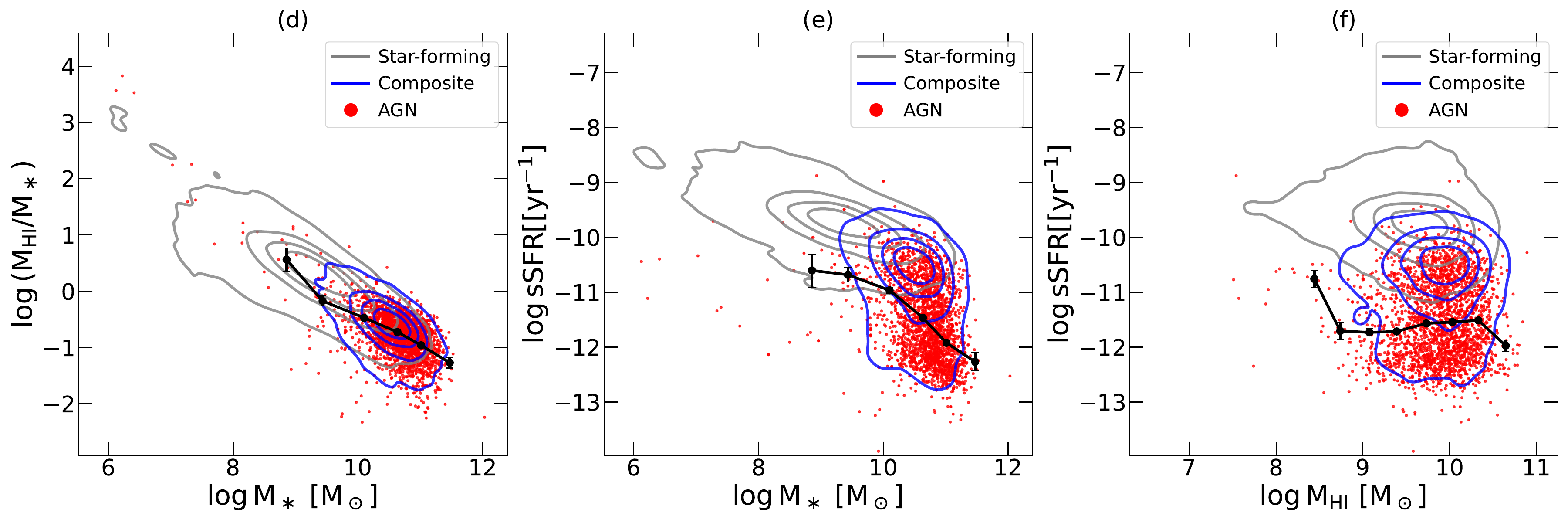}
\caption{Correlations among $\rm M_{\ast}$, $\rm M_{\hi}$, SFR and sSFR. (a): $\rm M_{\ast}-\rm M_{\hi}$; (b): $\rm M_{\ast}-$SFR;   
(c): $\rm M_{\hi}-$SFR; (d):  $\rm M_{\ast}-\rm M_{\hi}/M_{\ast}$; (e): $\rm M_{\ast}-$sSFR ; (f): $\rm M_{\hi}-$sSFR. 
The gray and blue contours represent the normalized density distributions of star-forming and composite galaxies. The contours represent the bivariate nuclear density estimation (KDE) for the control samples (e.g., star-forming and composite galaxies).
Respectively, with red points indicating individual AGN systems. The black solid line shows the median trend of the AGN systems, with dot symbols indicating the median values in individual bin.} 
\label{fig:HI/stellar}
\end{figure}

To investigate the impact of AGN activity on the cold gas reservoirs of galaxies, we compare the distributions of $\rm M_{\rm HI}$, $\rm M_\ast$, and  $\rm M_{\rm HI}/M_\ast$ among star-forming, composite, and AGN-hosting galaxies. Throughout the subsequent analysis, the AGN-hosting sample is defined to include both Seyferts and LINERs without further distinction. Furthermore, we examine the relations of these gas properties with the SFR and sSFR.  We are able to directly assess whether AGN hosts systematically deviate from normal star-forming galaxies in terms of their cold gas content and star formation efficiency. 
Figure~\ref{fig:HI/stellar} shows the relation between the $M_{\rm HI}$ and $M_\ast$ for star-forming (gray), composite (blue), and AGN (red) galaxies. 

In panel (a), the \hi mass increases with stellar mass, with the slope flattening at the high-mass end. 
The contour lines in each panel represent the bivariate Kernel Density Estimation (KDE) of the control galaxy populations. We set levels=5 for the KDE plots, where each contour represents a specific iso-proportion of the cumulative probability density. The outermost contour encompasses the vast majority of the sample distribution, while the innermost contour indicates the highest density region (the mode) of the population.
Star-forming galaxies follow a relatively tight correlation, whereas both composite and AGN-hosting galaxies are predominantly located at higher stellar masses ($\log M_\ast \gtrsim 10$) and exhibit systematically lower HI masses than star-forming galaxies of similar stellar mass. This is further reflected in panel (d), where the gas fraction ($M_{\rm HI}/M_\ast$) is anti-correlated with stellar mass. The optically selected AGN and composites lie systematically below the main $M_{\rm HI}$--$M_\ast$ relation defined by the star-forming population \citep{2003MNRAS.346.1055K}. This trend indicates that galaxies hosting active nuclei are generally more gas-poor, suggesting that AGN feedback may play a role in depleting or heating their cold gas reservoirs. Composite galaxies occupy an intermediate region, consistent with the picture that they are in a transitional phase from active star formation to quiescence.

Panel (b) illustrates the correlation between stellar mass and SFR, known as the star-forming main sequence (SFMS)  \citep{2004MNRAS.351.1151B,2009MNRAS.400..154B}.  Star-forming galaxies (gray contour) follow a tight, nearly linear correlation. In contrast, composite galaxies lie systematically below the SFMS, while AGN-host galaxies  exhibit significantly lower SFRs at fixed stellar mass, as indicated by the median solid line. As shown in \citet{2012ApJ...754L..29W}, this relation exhibits a nonlinear slope (SFR $\propto M_{\ast}^{0.6}$) with a scatter of 0.34 dex. The majority of AGN-hosting galaxies are located below the SFMS, a phenomenon also reported in recent studies \citep{2025MNRAS.539.3229G,2023ApJ...953...26L}. This suggests that AGN-hosting galaxies are typically more massive and have experienced suppressed or quenched star formation, possibly driven by AGN feedback \citep{2022A&A...665A.144M}. Panel (e) shows the sSFR--$M_\ast$ correlation, where sSFR decreases with increasing stellar mass \citep{2005MNRAS.358..363C}. Consistent with the SFR trends, AGN deviate from the SFMS, showing signs of suppressed star formation \citep{2015MNRAS.452.1841S,2007ApJS..173..267S}.

Panels (c) and (f) show how SFR and sSFR with \hi mass, respectively. SFR increases with \hi mass, consistent with the findings of ~\cite{2012ApJ...756..113H}. However, no clear global trend is discernible in the sSFR-$M_{\rm HI}$ relation, as indicated by the median solid line in panel (f). Instead, a population-dependent bifurcation is observed: star-forming galaxies are concentrated in the region of high sSFR at fixed gas mass, while AGN are concentrated in the lower-right region, indicating significantly reduced sSFR. Despite possessing comparable gas reservoirs to star-forming galaxies, AGN and composite galaxies exhibit much lower star formation efficiency. This provides further evidence that AGN feedback can suppress star formation even in galaxies with abundant cold gas.

\begin{figure}
\centering
\includegraphics[width=\textwidth]{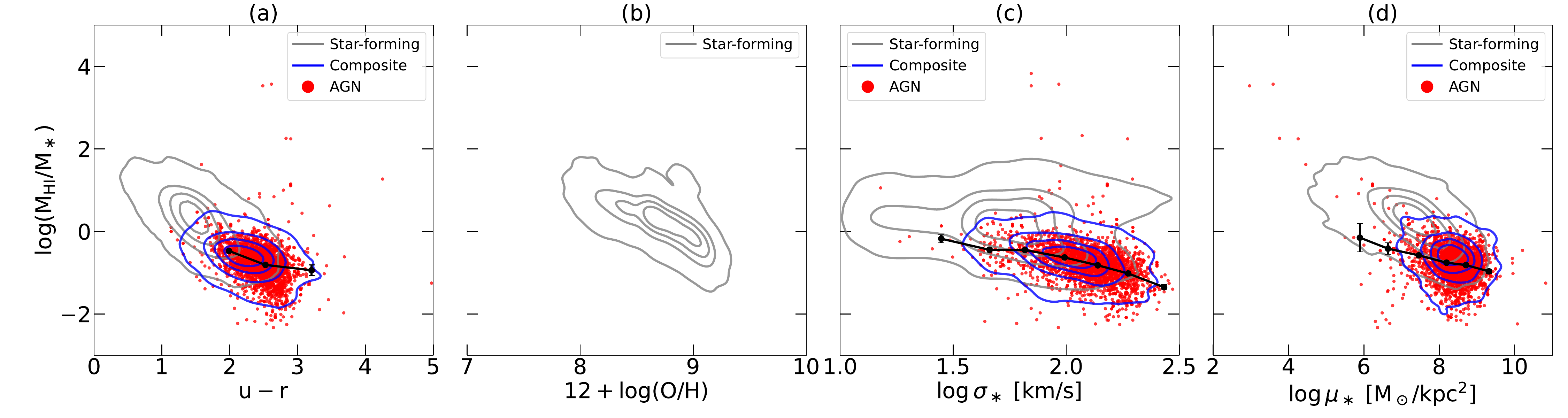}
\caption{\hi Gas fraction $\rm M_{\hi}$/$\rm M_{\ast}$ as function of $\rm u-r$ (panel a), $\rm log(O+H)+12$ (panel b), $\log \sigma_\ast$ (panel c), $\log \mu_\ast$ (panel d). The gray and blue contours represent the normalized density distributions of star-forming and composite galaxies, respectively, with red points indicating individual AGN systems. The black solid line shows the
median trend of the AGN systems, with dot symbols indicating the median values in individual bin.
}
\label{fig:HI_optical_par}
\end{figure}

\begin{figure}
\centering
\includegraphics[width=\textwidth]{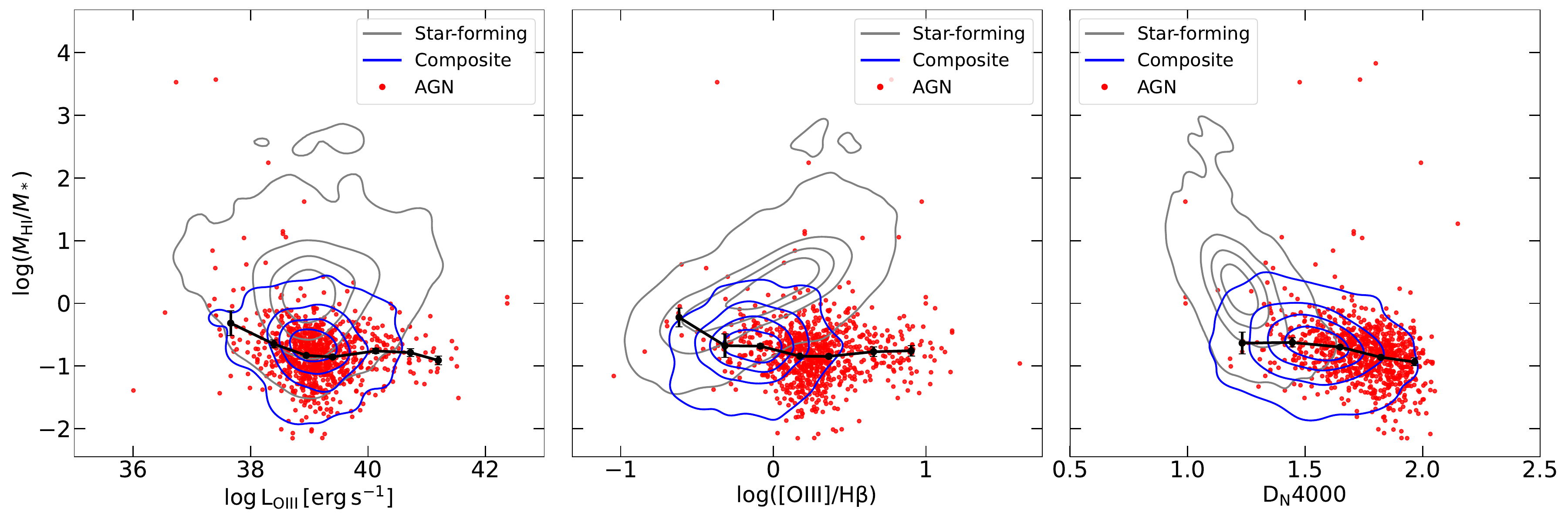}
\caption{Correlations between $\rm M_{\hi}$/$\rm M_{\ast}$ and $\log \rm L_{\rm OIII}$ (panel a), $\log([\rm OIII]/\rm H\beta)$ (panel b), $\rm D_{\rm N}4000$ (panel c). The gray and blue contours represent the normalized density distributions of star-forming and composite galaxies, respectively, with red points indicating individual AGN systems. The black solid line shows the median trend of the AGN systems, with dot symbols indicating the median values in individual bins.
}
\label{fig:HI_optical_par2}
\end{figure}

In Figure~\ref{fig:HI_optical_par}, we show the gas fraction $\rm M_{\hi}$/$\rm M_{\ast}$ as a function of various galaxy parameters, including the color $\rm u-r$, gas metallicity $\rm log(O+H)+12$, stellar velocity dispersion $\log \sigma_\ast$ and stellar mass surface density $\log \mu_\ast$. Panel (a) shows the relationship between gas content and optical color. Star-forming galaxies typically reside in the upper-left region (bluer, gas-rich), while AGN generally occupy the lower-right region (redder, gas-poor), with composite galaxies located in the intermediate transition zone. The
$\rm u-r $ colors of AGN-hosting galaxies are mainly concentrated between 2 and 3, indicating a pronounced clustering in this color range. This distribution is consistent with an evolutionary scenario in which galaxies become redder as their gas reservoirs are depleted or heated by feedback processes (e.g., supernova or AGN feedback) or environmental stripping, leading to the cessation of star formation. In panel (b), a negative correlation is expected between metallicity and gas fraction, where higher metallicity corresponds to a lower residual neutral hydrogen gas fraction. It should be noted, however, that strong-line metallicity calibrations cannot be reliably applied to AGNs and composite galaxies because their emission lines are contaminated by non-stellar ionizing radiation \citep{2003MNRAS.346.1055K}. In panels (c) and (d),
the HI gas fraction exhibits a gradual decreasing trend with increasing central stellar velocity dispersion ($\sigma_{\ast}$) and stellar surface density ($\mu_{
\ast
}$), indicating an anti-correlation with these parameters \citep{2005Natur.433..604D,2006MNRAS.365...11C,2009Natur.460..213C,2012ARA&A..50..455F,2024Natur.632.1009W}. This trend aligns with the findings of \cite{2010MNRAS.403..683C} and \cite{2015MNRAS.452.2479B}: galaxies with high $\sigma _{*}$ or $\mu_{*}$ have typically completed the evolutionary transition from gas-rich to gas-poor states. Star-forming galaxies predominantly occupy the low $\sigma _{*}$ and $\mu_{*}$ regime and are associated with relatively high HI gas content, whereas AGN and composite galaxies exhibit lower HI gas fractions.

These distributions indicate a coherent evolutionary sequence: galaxies transition from HI-rich, blue, low $\sigma _{*}$ and low $\mu_{*}$ star-forming systems through a composite phase toward HI-poor AGN-hosting galaxies characterized by dynamically hot and structurally compact stellar components\citep{2005Natur.433..604D,2006MNRAS.365...11C,2009Natur.460..213C,2012ARA&A..50..455F,2024Natur.632.1009W}. These results highlight the close connection between the HI gas, the buildup of central stellar structures, and the emergence of AGN activity.

In addition, we investigate how gas fraction and star formation activity correlate with key tracers of AGN activity: the [O\,\textsc{iii}] luminosity ($L_{\rm [OIII]}$), the ratio of $\rm [OIII]$-to-$\rm H \beta$ and the 4000\,\AA\ break strength ($D_N4000$). These tracers serve as independent indicators of nuclear activity and stellar population age, enabling us to assess whether stronger AGN activity or older stellar populations are linked to suppressed gas reservoirs and reduced star formation. This approach allows us to quantify the relative importance of AGN feedback compared to secular star formation processes in regulating the \hi content of galaxies. 
We have constructed a stellar mass-matched sample (see Section 4.3) to test whether the observed trends are driven by stellar mass differences between AGN-hosting and other galaxies.
In Fig \ref{fig:HI_optical_par2}  panel (a), 
star-forming and AGN-hosting galaxies largely overlap in $L_{[OIII]}$. The median values differ by 0.1 dex, from 39.0 for star-forming galaxies to 39.1 for AGN-hosting galaxies. The fraction of objects with $\rm log L[OIII] > 40.0$ is $10.1\% $ for AGN-hosting galaxies and $8.1\%$ for star-forming galaxies. Composite galaxies lie between these two populations. This figure shows that $L_{\rm [OIII]}$ does not show a clear dependence on the \hi gas fraction among the SF, composite, and AGN-hosting galaxies. 
Panel (b) shows that galaxies with higher $\rm [OIII]/\rm H \beta$ ratios, indicative of harder ionizing radiation fields, tend to have reduced \hi gas fractions. However, the median trend (solid line) is nearly flat, displaying only a weak and gradual decline, suggesting that the dependence is not very strong. The ratio of $\rm [OIII]/\rm H \beta$ could be an indicator of radiation field hardness \citep{2005ApJ...624..162H}. AGN-hosting galaxies are concentrated toward higher $\rm [OIII]$-to-$\rm H \beta$  values and lower \hi content, whereas star-forming galaxies dominate the low-ionization, \hi-rich region.
In panel (c), the \hi gas fraction decreases with increasing $\rm D_N4000$
, indicating a strong connection between atomic gas content and stellar population age. Star-forming galaxies, characterized by young stellar populations (low $\rm D_N4000$
), retain large \hi reservoirs, while AGN hosts preferentially reside in systems with older stellar populations and lower \hi gas fractions. Composite galaxies again occupy an intermediate locus.

For the AGN-hosting galaxies, we find no evidence that the gas fraction depends on the strength of the AGN activity: The \hi gas fraction cannot show a significant correlation with 
 $L_{\rm [OIII]}$ or the $\rm [OIII]/\rm H \beta$ ratio.

\subsection{The HI Gas in Satellite Galaxies}
We next examine how the environment affects the HI gas content of galaxies by comparing central and satellite systems within their host halos. Satellite galaxies, as they orbit within the gravitational potential of a more massive central host, are subject to environmental processes such as tidal stripping \citep{2006MNRAS.369.1021M,2024A&A...686A.184D,2020A&A...638A.133L,2022MNRAS.516.4293W}.
Besides, satellite galaxies also experience ram pressure stripping, resulting from their high-velocity interaction with the intragroup medium (IGM) \citep{1972ApJ...176....1G,2006MNRAS.370..453R,2006PASP..118..517B}.
These mechanisms can efficiently remove cold gas from satellites, potentially leading to a significant depletion of their HI reservoirs compared to central galaxies. To quantify this environmental impact, we analyze the HI gas fraction as a function of group-centric radius. We define $R_{180}$ as the radius within which the average density is 180 times the critical density of the universe; this serves as a proxy for the halo boundary. This radius can be derived from the halo mass via \cite{2007ApJ...671..153Y}:

\begin{equation}
R_{180} = 0.781\, h^{-1} \text{ Mpc} \left( \frac{M_h}{\Omega_m \times 10^{14} \, h^{-1} \, \text{M}_{\odot}} \right)^{1/3} (1 + z_{\text{group}})^{-1},
\label{eq:R180}
\end{equation}
where $M_h$ is the halo mass of the group, $\Omega_m$ is the present-day matter density parameter, and $z_{\text{group}}$ is the group redshift. The factor $(1+z_{\text{group}})^{-1}$ accounts for the conversion from comoving to physical radius.

\begin{figure}
\centering
\includegraphics[width=\textwidth]{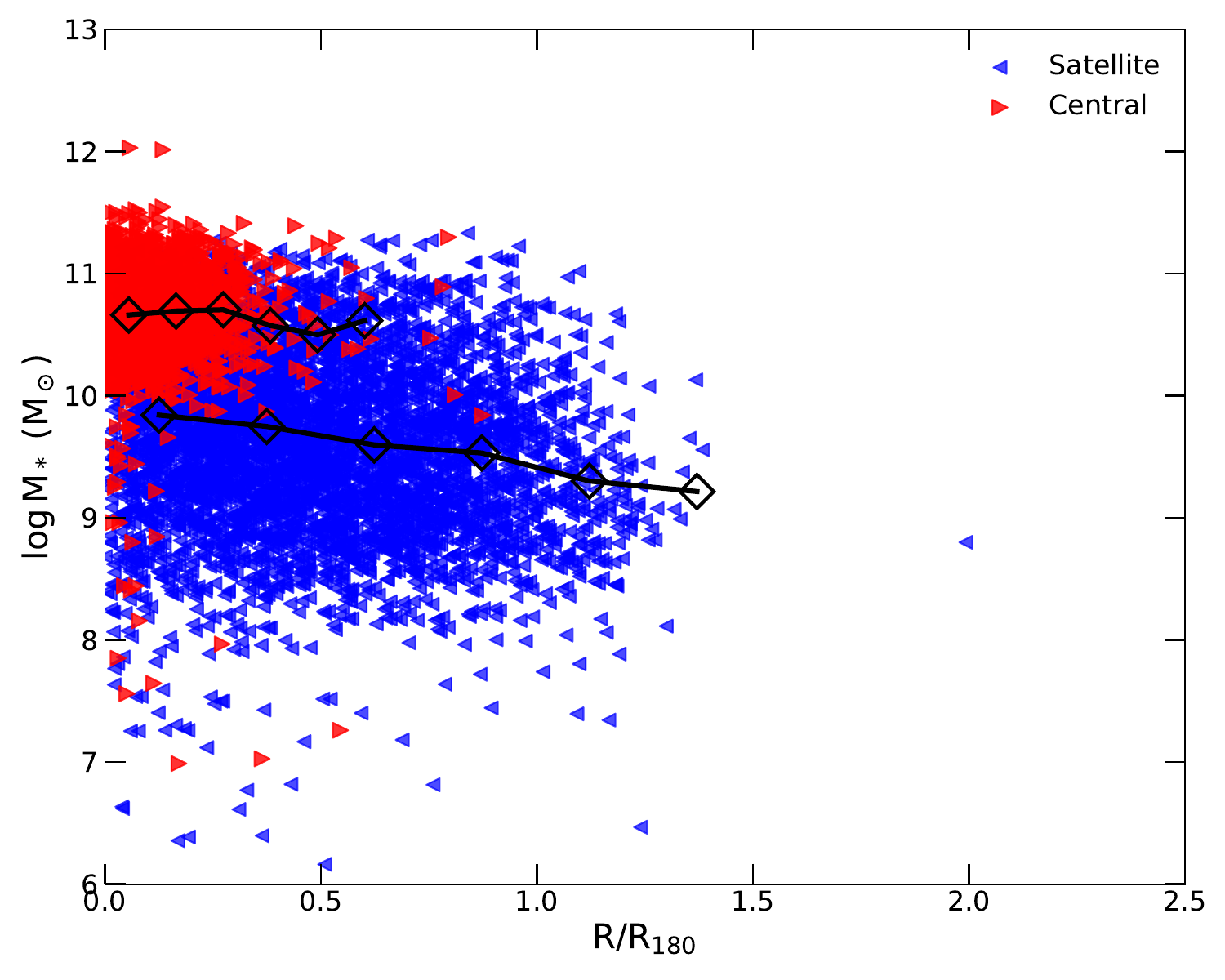}
\caption{The correlations between $\rm M_{\ast}$ and normalised projected group-centric radius for all galaxies in our sample. The red right-triangles blue represent the central galaxies and blue left-triangles represent the satellite galaxies, respectively.
The black solid line respectively shows the median
trend of the Central and Satellites, with open diamond symbols indicating the median values in individual bin.}
\label{fig:R_R180_M*_par}
\end{figure}

We selected all group central galaxies and satellite galaxies from our sample. Using the $R/R_{180}$ values calculated from Equation ~\ref{eq:R180}, we show the distribution of stellar mass as a function of group-centric radius for both centrals and satellites in Figure ~\ref{fig:R_R180_M*_par}. Central galaxies (red right-pointing triangles) are, as expected, concentrated near the potential well of the groups ($R/R{180} \approx 0$) and are preferentially found at higher stellar masses. In contrast, satellite galaxies (blue left-pointing triangles) exhibit a broad spatial distribution extending to $R/R_{180} \approx 1.5$ and span a wide range of stellar masses, from $\sim 10^{6.5}$ to $\rm 10^{11.5}M_{\odot}$. This comprehensive coverage enables a robust statistical comparison of \hi gas stripping and star formation quenching across different group environments.

To investigate the physical mechanisms linking \hi gas to group environment, we categorize the satellites based on their host halo mass.  To minimize the impact of stellar mass, we further restrict the satellite sample to a specific mass range with stellar masses $\rm M_\ast$ between $10^{10}$ and $10^{11.5}\,\rm M_{\odot}$. 
We analyzed the relations between the \hi content of galaxies and their group-centric radius, including the \hi mass, \hi-to-stellar mass ratio, and star formation rate. It is noteworthy that the \hi mass, \hi gas fraction, and \hi deficiency of a galaxy are fundamentally dependent on its stellar mass. 
 The satellite sample is subdivided according to the host group halo mass, using a threshold of $M_{\text{halo}} = 10^{13.5}\,M_{\odot}$. This yields two subsamples representing high-mass ($M_{\text{halo}} > 10^{13.5}\,M_{\odot}$) and low-mass ($M_{\text{halo}} \leq 10^{13.5}\,M_{\odot}$) environment \citep{2016ApJ...824..124E}.
 To avoid the effect from the stellar mass, we also have constructed a stellar mass-matched sample. Specifically, we performed a one-to-one matching procedure without replacement.
For each galaxy in the sample, we selected at most one control galaxy with similar stellar mass, requiring $\rm \Delta log M_{\ast} < 0.1$ dex and $\Delta z < 0.01$.

\begin{figure}
\centering
\includegraphics[width=\textwidth]{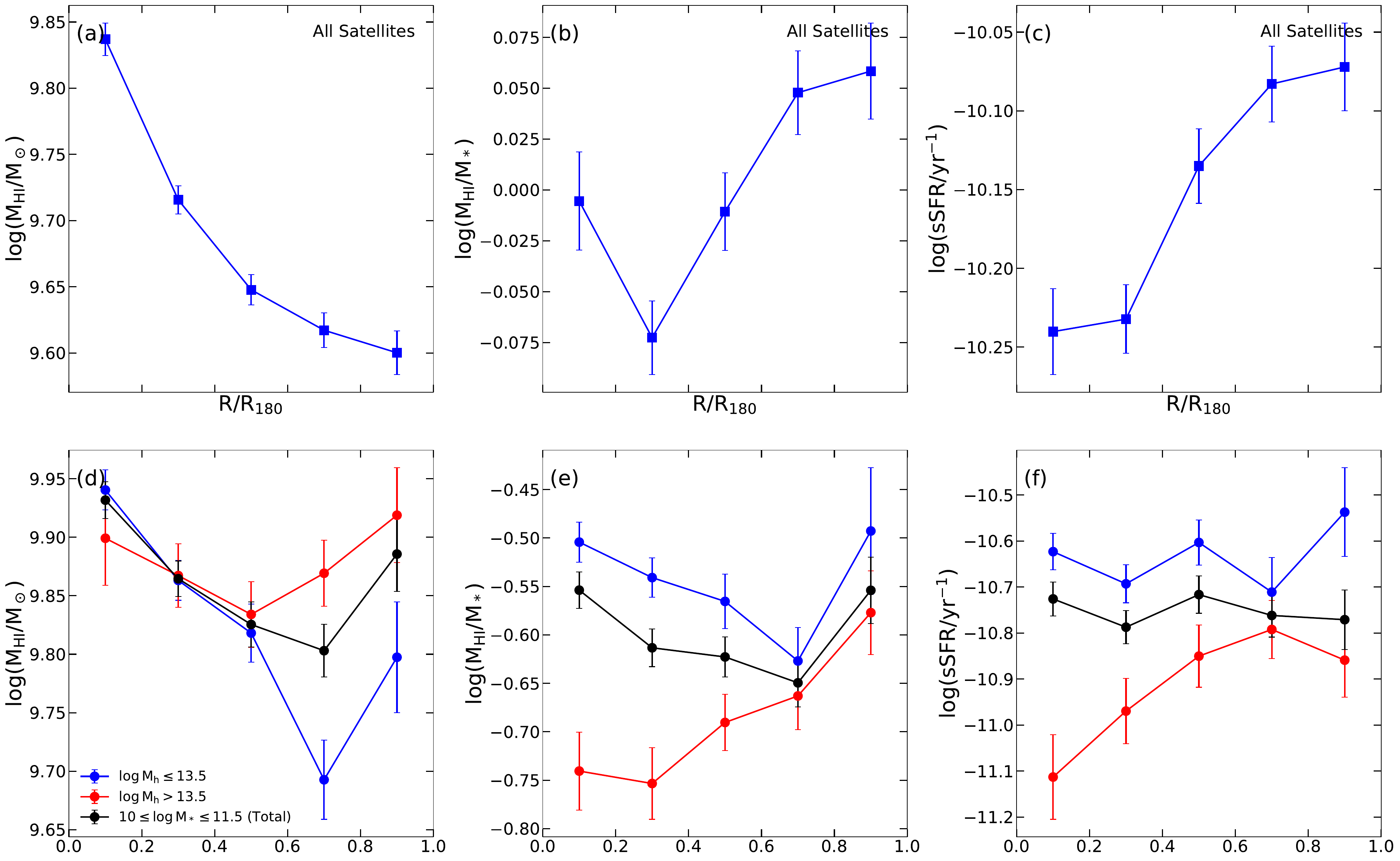}
\caption{Group-centric radius distribution of gas and star formation properties for satellites. The satellite sample is subdivided according to the
host group halo mass, using a threshold of $\log M_h/M_{\odot} =  13.5$. 
Top panels (a–c): Profiles for the entire satellite sample without any stellar mass cuts. 
Bottom panels (d–f): Profiles for satellites with stellar masses $10 \le \log(M_*/M_\odot) \le 11.5$, subdivided by host halo mass into low-mass ($\log(M_h/M_\odot) \le 13.5$, blue circles) and high-mass ($\log(M_h/M_\odot) > 13.5$, red circles) groups. The black circles represent the combined total within this stellar mass range. The error bars are estimated via the bootstrap method.
Some inconsistencies in the innermost bins (e.g., the non-monotonic behavior in low-mass halos) are considered to be the potential impact of beam confusion from the central galaxies.
}
\label{fig:R_R180_HI_divide_13.5_par}

\end{figure}

Top panels in Figure \ref{fig:R_R180_HI_divide_13.5_par} illustrate the statistical trends for the all selected satellite galaxy sample as a collective whole, without any constraints on stellar mass or segregation by host halo mass. The \hi mass exhibits a clear decline as a function of increasing group-centric radius. In the region where $R/R_{180} > 0.3$, the gas fraction decreases steadily with decreasing radius. However, a higher gas fraction is observed in the innermost region ($R/R_{180} \approx 0.1$). The sSFR of satellite galaxies outside is significantly stronger than that in the central regions.

The stellar-mass-matched results shown in the bottom panels are used as a robustness check to reduce the possible influence of different stellar-mass distributions among the compared satellite samples. In this test, the median \hi gas content is recalculated in bins of $R/R_{180}$ after matching the stellar-mass distributions of the satellite galaxies. 
Bottom panels in Figure \ref{fig:R_R180_HI_divide_13.5_par} illustrate the radial trends of \hi gas content and star formation activity for satellite galaxies with stellar masses $10 \le \log(M_*/M_\odot) \le 11.5$. Blue circles represent the satellite galaxies residing in low-mass halo environments ($\log M_{halo} \leq 13.5M_\odot$), while red circles represent those in high-mass halo environments ($\log M_{halo} > 13.5M_\odot$). The black circles provide a baseline representing the combined total within this stellar mass range.
In panel (d), the satellite galaxies in low-mass halos (blue circles) exhibit a distinct and clear downward trend within the range of $0.1 < R/R_{180} < 0.7$. The average \hi mass decreases from approximately $9.94$ to $9.69$. In the outermost region ($R/R_{180} \approx 0.9$), the \hi mass shows a slight increase recovery. Panel (e) illustrates the \hi gas fraction relative to the stellar mass. The trend is highly similar to that in Panel (d). 
In contrast, for satellite galaxies in high-mass halos (red circles), the average \hi mass in panel(d) reaches a local minimum at $R/R_{180} \approx 0.5$ before increasing toward both the center and the outer regions. However, when normalized by stellar mass in panel (e), red circles show a clear tendency, with the gas fraction declining steadily as the radius decreases. This reflects more intense ram-pressure stripping or tidal effects in high-mass halos.
This may suggest that in low-mass halos (e.g., galaxy groups), environmental effects are weaker than in high-mass halos (e.g., galaxy clusters). 
In Panel (f), this panel shows the star formation rate of the satellite galaxies. The blue curve fluctuates significantly across the radial range, with no evident increasing or decreasing trend. We can find that these galaxies maintain a certain level of star formation activity even in the central regions of low-mass halos. However, the red curve displays a significant tendency, with a marked reduction in sSFR for galaxies closer to the center, indicating strong environmental quenching in massive halos. 
A comparative analysis of blue and red profiles may reveal that the radial profiles of gas fraction and sSFR are less severely impacted in the low-mass regime. This suggests that environmental mechanisms, such as tidal stripping, are notably less efficient in low-mass halos (groups) compared to their high-mass counterparts (clusters).
The black circles, representing the combined satellite population ($10 \le \log(M_*/M_\odot) \le 11.5$), exhibit radial trends that closely resemble those of satellites in low-mass halos. This suggests that the statistical properties of the overall satellite population are dominated by galaxies residing in lower-mass group environments.

The resulting radial behavior remains qualitatively similar to that obtained from galaxies without stellar-mass matching. We can find that the number of high-stellar-mass galaxies in our selected sample is relatively small. This suggests that, although stellar mass can affect the absolute \hi gas content, the observed dependence on $R/R_{180}$ is unlikely to be driven solely by stellar-mass differences, and environmental processes may play an important role.

The observational work of \cite{2021MNRAS.507.5580H} has confirmed a decreasing trend of \hi mass for satellite galaxies from the outside to the halo centre. 
To address the discrepancies in the analysis, a observational constraint in this study is the spatial resolution of the \hi surveys. Both FAST and ALFALFA have a beam size (FWHM) of larger than 2.95 arcmin, which is insufficient to spatially resolve satellite galaxies that are in close proximity to the group center or their primary central galaxies. At a small group-centric radius (low $R/R_{180}$), the angular separation between a satellite and its neighbors often falls within a single beam. This results in beam confusion, where the detected \hi signal for a target satellite may be contaminated by gas emissions from the central galaxy or other nearby companions. Consequently, this leads to an overestimation of the \hi mass and gas fraction at low $R/R_{180}$. This rebound effect is more pronounced in low-mass halos. This is due to the smaller physical separation between the satellite and the central galaxy in these systems. This observational bias may explain the lack of a sharp radial decrease in \hi content near the group center.

\subsection{AGN feedback and Tidal forces in Satellite Galaxies}

In our sample, we identify various galaxy populations, including AGN-hosting, star-forming, and composite galaxies, among both satellite and central galaxies. A key objective of this study is to analyze the respective roles of environmental tidal forces and  AGN feedback mechanisms to determine which process plays a more dominant role in regulating the \hi gas. While these mechanisms are often mass-dependent, our current scatter and contour analysis does not yet account for the differentiation across different stellar mass ($M_*$) regimes, which remains a critical factor for a comprehensive understanding of these processes.

\begin{figure}
\centering

\includegraphics[width=\textwidth]{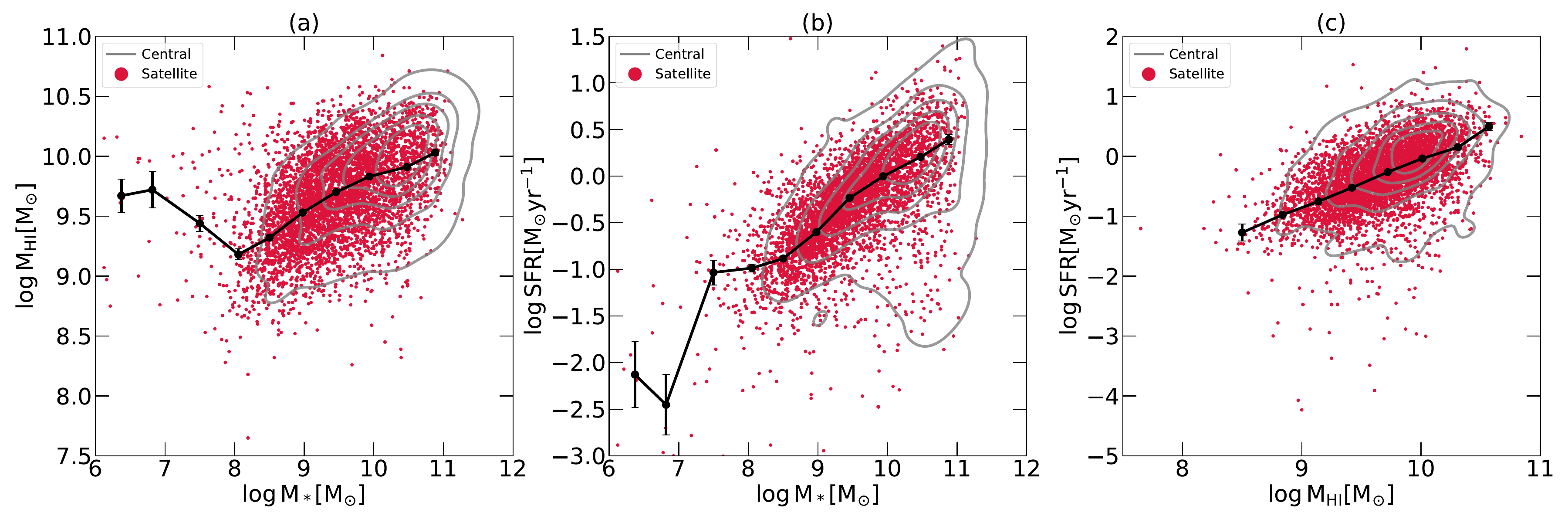}
\includegraphics[width=\textwidth]{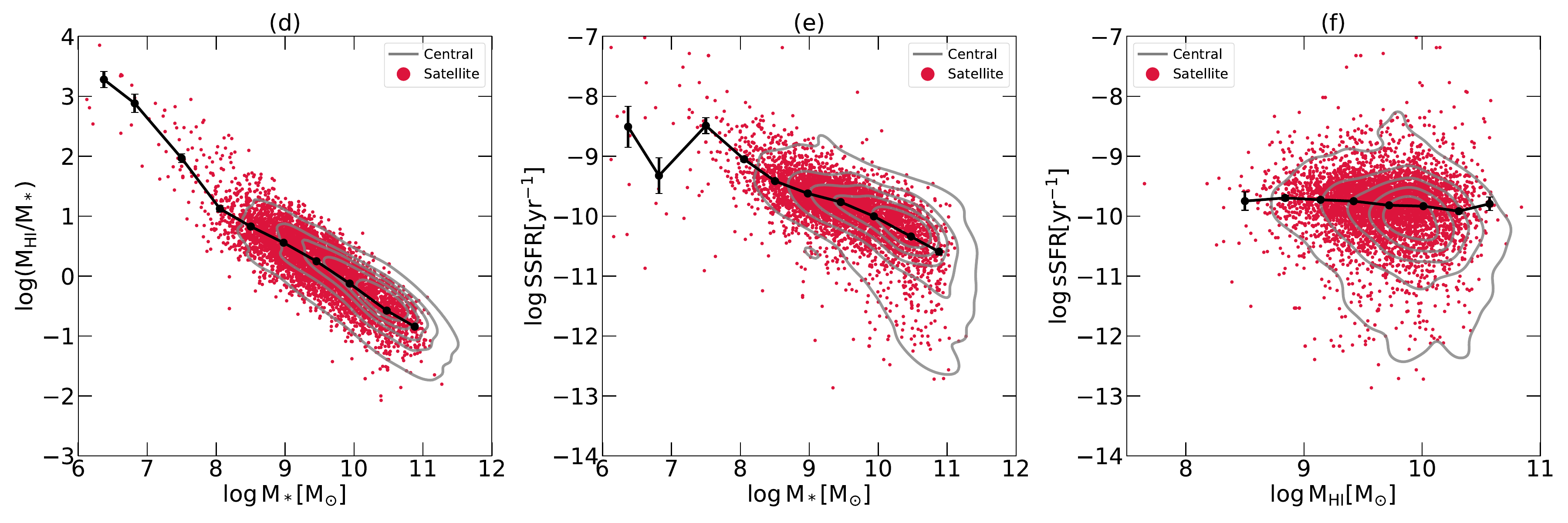}
\caption{The correlations among $M_\ast$, $M_{HI}$, SFR and sSFR. The red dots represent the satellite galaxies, and the gray contours represent the central galaxies, respectively.
The black solid line shows the
median trend of the Satellites, with dot symbols indicating the median values in individual bin.}
\label{fig:no_agn_cen_satellites_hi_par}
\end{figure}

\begin{figure*}
\centering

\includegraphics[width=\textwidth]{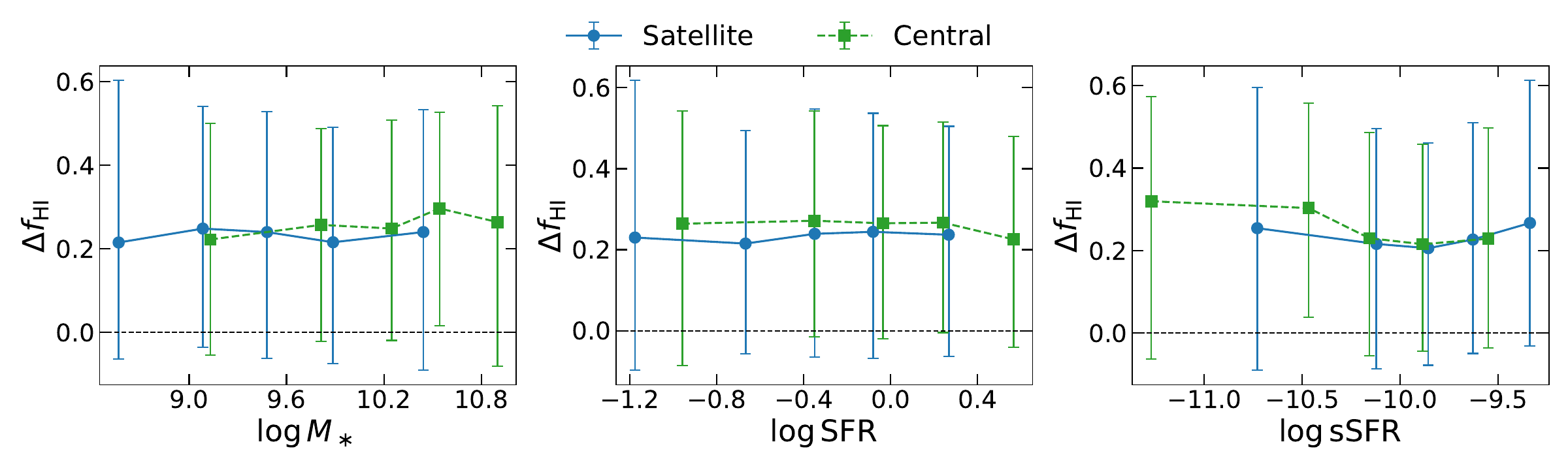}
\caption{ Median $\rm \Delta f_{HI}$ of central and satellite galaxies against stellar mass, SFR, sSFR. Central and satellite galaxies are shown using green and blue,
respectively.
}
\label{fig:delta_log_fhi}
\end{figure*}

In Figure ~\ref{fig:no_agn_cen_satellites_hi_par}, we display the physical parameter distributions for central and satellite populations, excluding those hosting an AGN. Isolated central galaxies are also omitted from the primary analysis. Our sample consists of 5,284 satellite galaxies and 2,384 group-central galaxies; for comparison, there are 16,327 galaxies identified as isolated centrals in the sample.  We examine the distributions of \hi mass, stellar mass, gas fraction, SFR, and sSFR for our sample. The resulting scatter plots and contours reveal no statistically significant differences between the satellite and central galaxy populations. Therefore, we use the HI gas-fraction offset \citep{1983AJ.....88..881G,2010MNRAS.403..683C,2014MNRAS.444..667D,2026ApJ...997..254Y}, $\Delta f_{HI}$, to quantify their relative HI content in Figure \ref{fig:delta_log_fhi}. It is also worth noting that satellite galaxies typically have lower stellar masses and \hi gas content.

\begin{figure}
\centering

\includegraphics[width=\textwidth]{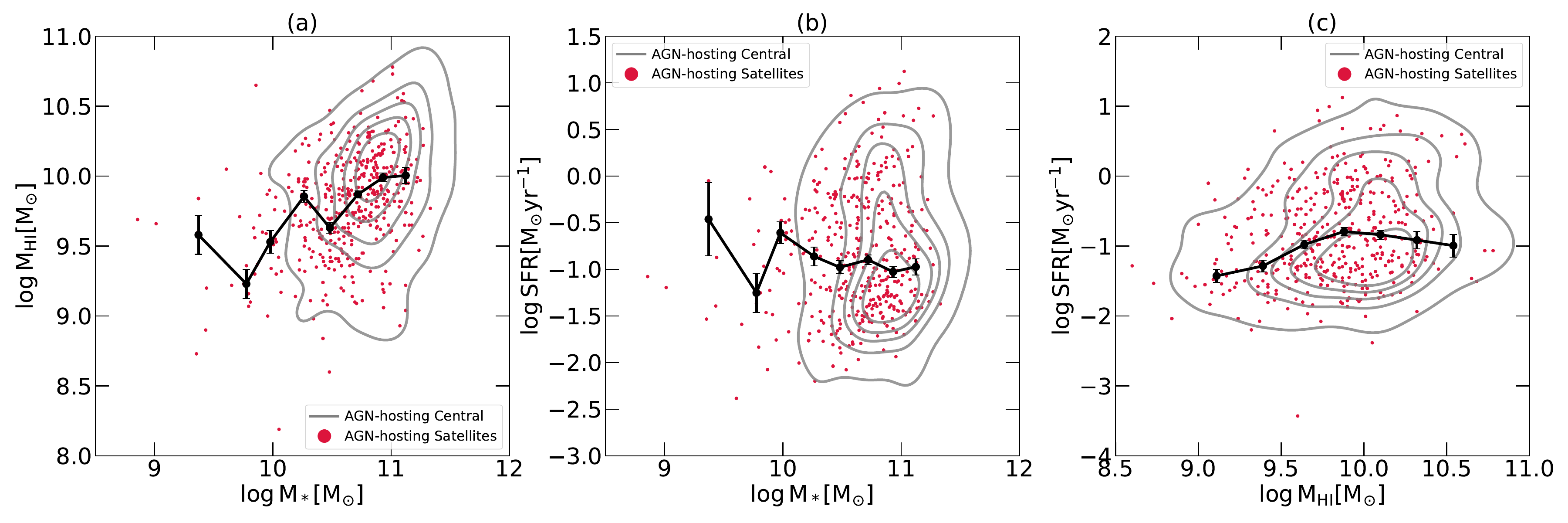}
\includegraphics[width=\textwidth]{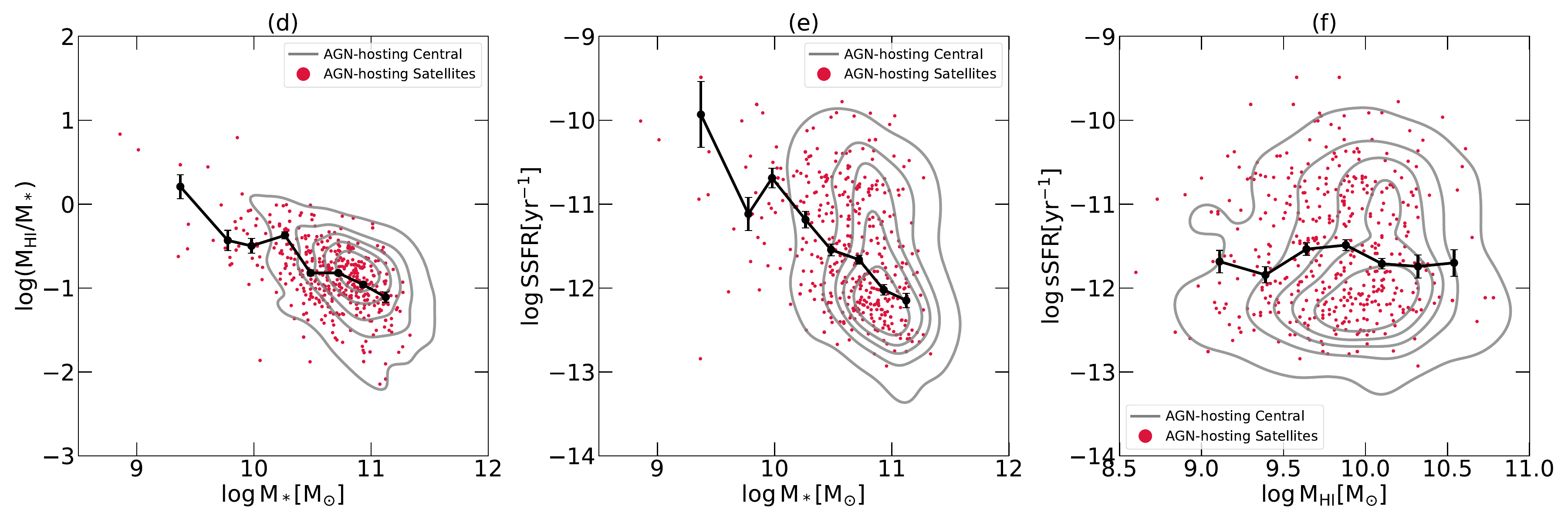}
\caption{The correlations among $\rm M_\ast$, $\rm M_{HI}$, SFR and sSFR. The red dots represent the AGN-hosting satellite galaxies, and the gray contours represent the AGN-hosting central galaxies, respectively.
The black solid line shows the
median trend of the AGN-hosting satellites, with dot symbols indicating the median values in individual bin.}
\label{fig:agn_cen_satellites_hi_par}
\end{figure}

We also display the physical parameter distributions for the AGN-hosting subsample in Figure ~\ref{fig:agn_cen_satellites_hi_par}, focusing on central and satellite populations while omitting isolated centrals. This subset includes 503 satellites and 891 group central galaxies, compared to a population of 1,129 isolated centrals. 
As shown in Figure ~\ref{fig:HI/stellar}, We find that AGN-hosting galaxies have higher stellar masses ($M_\ast$) and lower \hi gas content compared to the normal galaxies without AGN. Furthermore, no distinct differences are identified between AGN-hosting centrals and AGN-hosting satellites in their physical parameter distributions. This lack of distinction highlights a key limitation: simply examining the global scatter and contour distributions does not yield a definitive separation of their physical states, as it cannot distinguish between environmental quenching and internal feedback mechanisms. To break this degeneracy, we adopt a radial trend analysis in the previous section, examining how physical properties vary as a function of $R/R_{180}$.

\begin{figure}
\centering

\includegraphics[width=\textwidth]{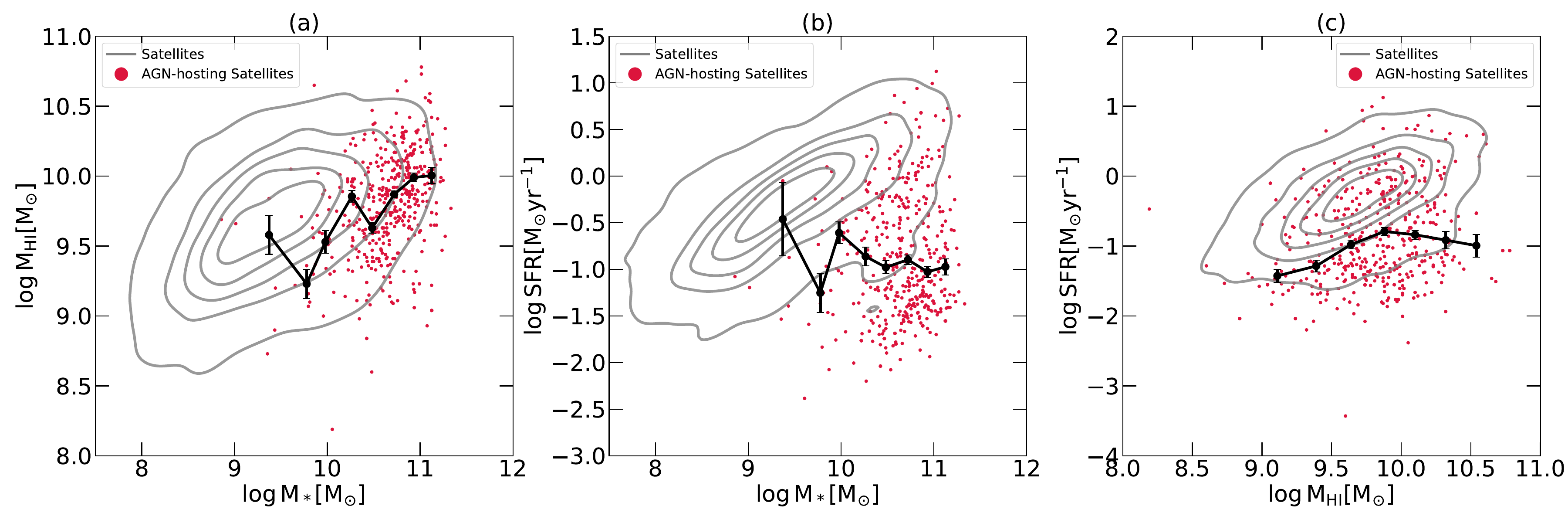}
\includegraphics[width=\textwidth]{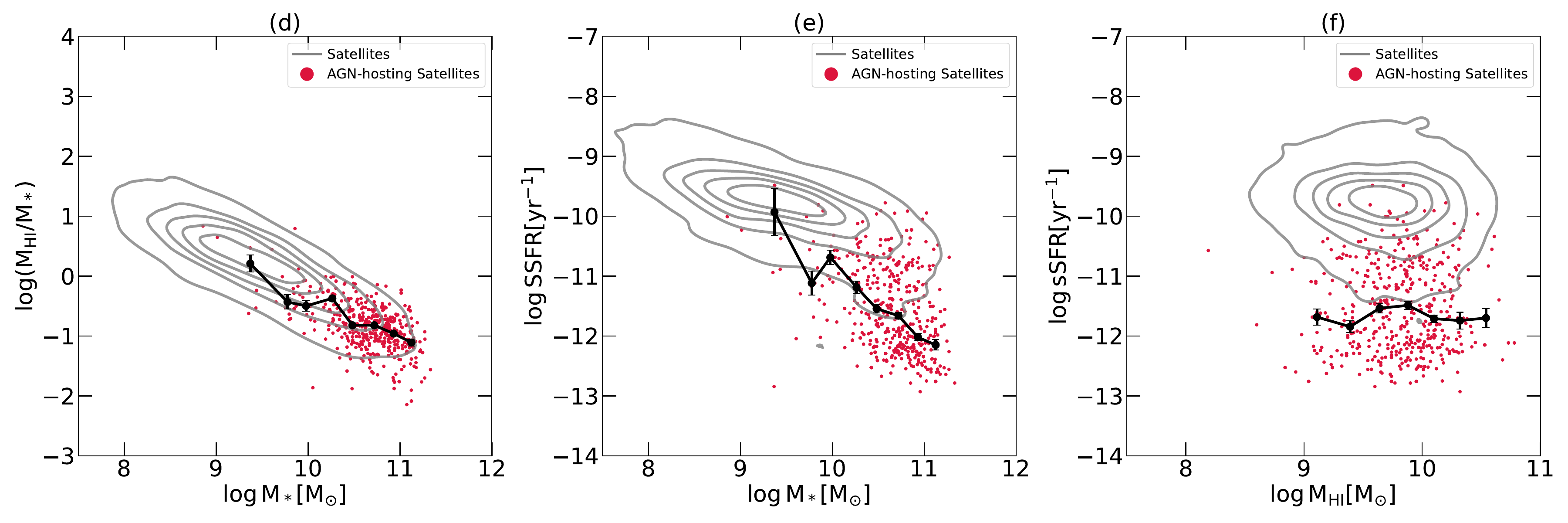}
\caption{The correlations among $\rm M_\ast$, $M\rm _{HI}$, SFR and sSFR. The red dots represent the AGN-hosting satellite galaxies, and gary contours represent the normal satellite galaxies without AGN, respectively. The black solid line shows the
median trend of the AGN systems, with dot symbols indicating the median values in individual bin.
}
\label{fig:agn_satellites_hi_par}
\end{figure}

In Figure ~\ref{fig:agn_satellites_hi_par}, we show the 
physical parameter distributions for the satellites, focusing on AGN-hosting and normal populations withou AGN. This subset includes 503 AGN-hosting satellite galaxies and 5284  normal satellite galaxies. We find that the parameter distributions for AGN-hosting satellites and normal satellites exhibit a trend consistent with that observed for SF, composite, and AGN galaxies in Figure ~\ref{fig:HI/stellar}. For AGN-hosting satellites, AGN feedback may play a significant role in depleting or heating their cold gas reservoirs. However, we cannot definitively determine whether AGN feedback precedes environmental effects like tidal stripping or if they operate concurrently.

To probe environmental effects more directly, we examine these physical properties as a function of normalized group-centric radius ($R/R_{180}$). To control for stellar mass, we restrict the analysis to satellite galaxies in a narrow stellar mass range ($10^{10} < M_\ast < 10^{11.5}\rm M_{\odot}$). Furthermore, we constructed a stellar mass-matched sample to avoid the effects of stellar mass.

\begin{figure*}
\centering

\includegraphics[width=\textwidth]{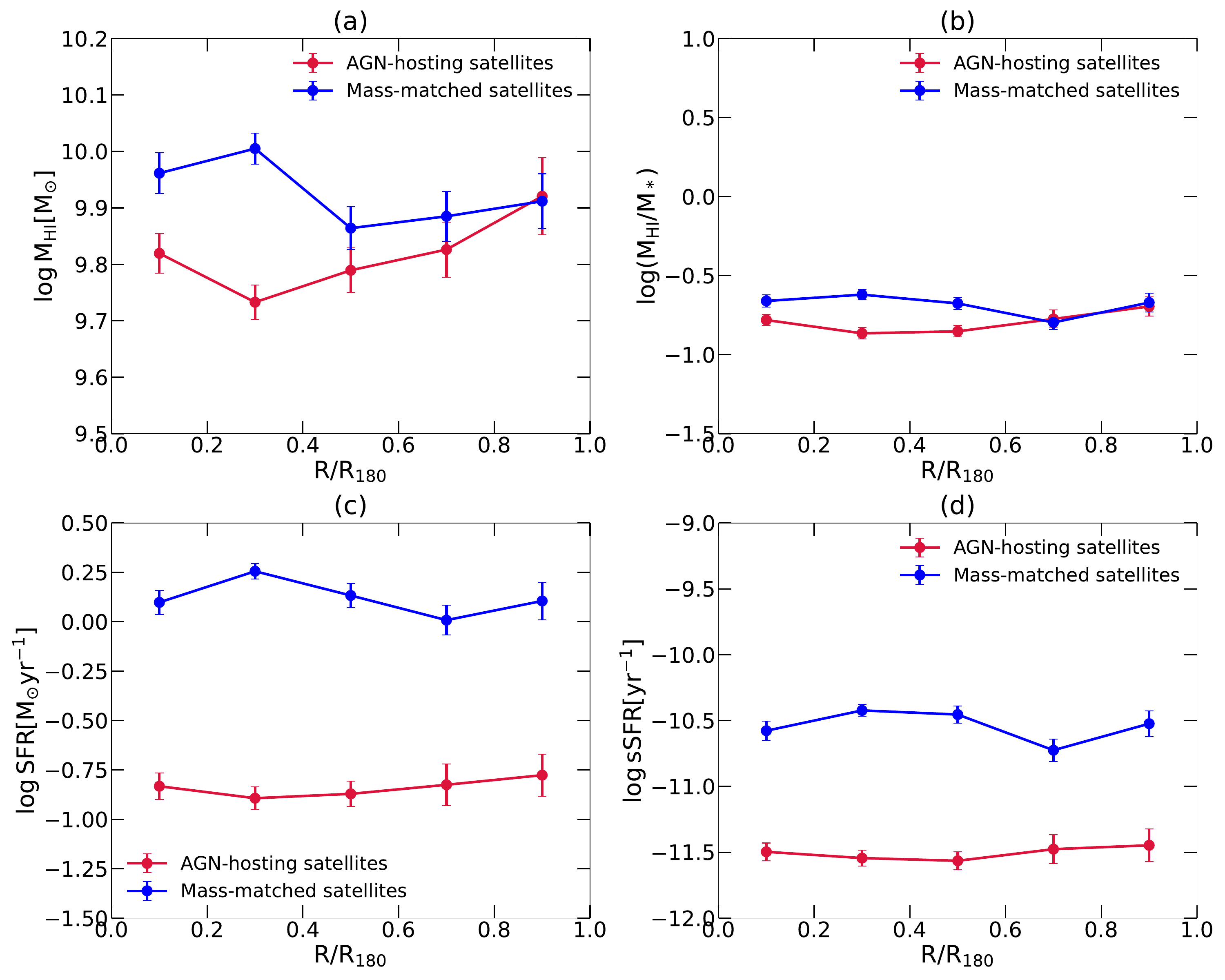}
\caption{We show the correlations between $\rm M_{HI}$, $\rm M_{\ast}$, SFR, sSFR, and normalised projected group-centric radius for all galaxies in our sample. The blue error bars represent the normal satellite galaxies, and the red error bars represent the AGN-hosting satellite galaxies, respectively. The data points represent the mean values in each radial bin, with
error bars denoting the 1$\sigma$ uncertainty derived from bootstrap resamplings.
}
\label{fig:agn_satellites_hi_R_R180_par}
\end{figure*}

Figure ~\ref{fig:agn_satellites_hi_R_R180_par} shows that the \hi gas fraction of AGN-hosting satellites are significantly lower than those of normal satellites without AGN at low group-centric radius($R/R_{180} \lesssim 0.7$)\citep{2011MNRAS.416.1739F,2012ApJ...758...73S}. This indicates a direct correlation between AGN activity and a more gas-poor state. At large group-centric radii ($R/R_{180} \gtrsim 0.7$), both AGN-hosting and mass-matched control satellites exhibit similar HI gas fractions, suggesting that galaxies in the group outskirts have not yet experienced significant environmental stripping.
Panels (c) and (d) show that AGN-hosting satellites exhibit a strong suppression of star formation. Their sSFR is generally about one order of magnitude lower than that of normal satellites. This suggests that AGN feedback may have effectively quenched these galaxies, moving them from the star-forming main sequence into the green valley or the quiescent regime \citep{2003MNRAS.346.1055K,2014SerAJ.189....1S}. 
We find the substantial gap between the red and blue curves at $R/R_{180} \approx 0.9$, which means near the outer region of the group halo. It implies that these AGN satellites have undergone gas expulsion or star formation suppression via AGN feedback before deep entry into the group center \citep{2012ARA&A..50..455F} and being influenced by tidal stripping or ram-pressure stripping. In panels (b), (c), and (d), as $R/R_{180}$ decreases (closer to the group center), both curves exhibit relatively flat trends without an obvious tendency. 
Environmental processes such as tidal interactions and ram-pressure stripping can also remove gas from satellites.
However, the relatively flat radial trends suggest that for this specific sample, internal AGN feedback may play a more dominant role in gas depletion and quenching than environmental factors such as tidal forces or ram-pressure stripping from the central galaxy. 
As shown in panel (a), we find that AGN-hosting satellites exhibit lower HI gas content compared to normal satellites.
Additionally, the apparent increase in \hi mass and gas fraction at $R/R_{180} < 0.3$ is likely attributable to beam confusion effects \citep{2012MNRAS.427.2841F,2021MNRAS.507.5580H} arising from the limited spatial resolution of the FAST and ALFALFA surveys.

\begin{figure*}
\includegraphics[width=\textwidth]{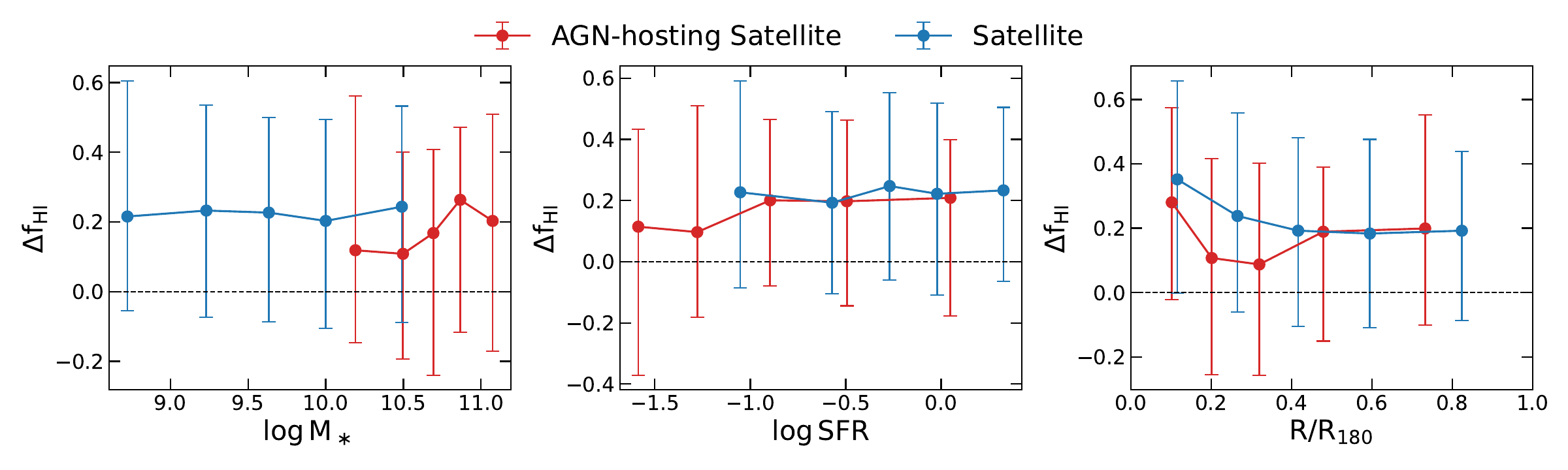}
\caption{ Median $\rm \Delta f_{HI}$ of AGN-hosting satellite and normal satellite galaxies against stellar mass, SFR, normalised projected group-centric
radius. AGN-hosting satellite and normal satellite galaxies are shown using red and blue,
respectively.
}
\label{fig:delta_log_fhi_agn}
\end{figure*}

Figure ~\ref{fig:delta_log_fhi_agn} shows that $\rm \Delta f_{HI}$ of central and satellite galaxies against stellar mass, SFR, normalised projected group-centric
radius. AGN-hosting satellite galaxies have higher stellar masses, lower SFRs, and lower $\rm \Delta f_{HI}$ than satellite galaxies without AGN. When examined as a function of the group-centric distance, R/R180, AGN-hosting satellites show systematically lower  $\rm \Delta f_{HI}$ at smaller R/R180. Moreover, their  $\rm \Delta f_{HI}$ values are lower than those of normal satellites at comparable group-centric distances. This trend suggests that AGN feedback may contribute significantly to the reduction of \hi gas content in AGN-hosting satellite galaxies.

\section{Summary and discussion}
\label{sect:summary and discussion}

In this study, we investigated the relative importance of internal AGN feedback versus environmental processes in regulating the cold gas content and star formation of galaxies. By combining \hi data from the ALFALFA and FASHI surveys with optical spectroscopy from SDSS DR7/DR8 and the DESI, we compare the physical properties of AGN-hosting galaxies in group environments against their normal galaxies without AGN.

Using the $L_{\text{[OIII]}}$ luminosity and the [OIII]/H$\beta$ ratio, and the $D_N4000$ index as indicators of AGN activity and stellar population age, we find no evidence that the gas mass fraction depends on the intensity of AGN activity. We also find that AGN-hosting galaxies experience a dramatic suppression in both star formation rate (SFR) and specific SFR (sSFR), approximately one order of magnitude lower than the star-forming population.  This quenching is accompanied by significatnly lower \hi gas fractions, observed across all group-centric radii.

While global scaling relations (e.g., $\rm M_{HI}$ vs $\rm M_*$) show significant overlap between central and satellite populations, the \hi gas-fraction offset and the radial trends reveal a critical distinction. A persistent and substantial gap in SFR persists between AGN-hosting satellites and normal satellites without AGN even at the virial radius ($R/R_{180} \approx 1$). This indicates that the depletion of cold gas and the suppression of star formation occur prior to the galaxy experiencing the full force of the group environment (e.g., tidal interactions or ram-pressure stripping). The relatively flat radial profiles of AGN-hosting galaxies suggest that their evolution is governed primarily by internal feedback rather than external processes driven by the central group galaxy.

We note an apparent increase in \hi mass and gas fraction at $R/R_{180}<0.3$ for AGN-hosting satellites, which is likely an artifact due to beam confusion stemming from the limited spatial resolution of single-dish surveys like FAST and ALFALFA. Notwithstanding this limitation, the overarching trend is robust: for the AGN-hosting population, internal feedback acts as the primary quenching mechanism, initiating gas heating or expulsion before the galaxy is subsequently affected by the secondary process of environmental stripping within the group halo.



\begin{acknowledgements}
This work is supported by the National SKA Program of China (Nos. 2022SKA0110100 and 2022SKA0110101), the National Natural Science Foundation of China (NSFC) International (Regional) Cooperation and Exchange Project (No. 12361141814), the NSFC innovation group grant 12421003, the National Key Research and Development Program of China (No. 2023YFA1607904), the National Natural Science Foundation of China (Nos. 12533001, 12473091), the Specialized Research Fund for State Key Laboratory of Radio Astronomy and Technology, the National Astronomical Observatories, Chinese Academy of Science (Nos. E5ZB0901, E4ZR0510), the Beijing Municipal Natural Science Foundation (grant no. 1242032), the Youth Innovation Promotion Association of the Chinese Academy of Sciences (No. 2022056), and the China Manned Space Program with grant Nos. CMS-CSST-2025-A07,  CMS-CSST-2021-B01, and CMS-CSST-2021-A01.
\end{acknowledgements}

\bibliographystyle{raa} 
\bibliography{bibtex}

\label{lastpage}

\end{document}